# Trends in EHR Satisfaction and Interoperability Among Family Physicians: A Five-Year Analysis of the ABFM Continuous Certification Questionnaire, 2022–2026

**Authors:** Carl Y. Zhang, MS; Nathaniel Hendrix, PharmD, PhD; Robert L. Phillips, MD, MSPH; Andrew W. Bazemore, MD, MPH

**Corresponding author:**

Nathaniel Hendrix, PharmD, PhD
Center for Professionalism and Value in Health Care, American Board of Family Medicine
1016 16th St NW Ste 800
Washington, DC 20036
United States of America
nhendrix@theabfm.org
+1 (202) 525-6918

**Affiliations:** All authors: Center for Professionalism and Value in Health Care, American Board of Family Medicine, Washington, DC, USA.

**Keywords:** Electronic Health Records; Health Information Interoperability; Health Information Exchange; Physicians, Family; Personal Satisfaction

**Abstract**

**Objective:** To assess trends in family physicians' (FPs) electronic health record (EHR) satisfaction and their experience of interoperability across organizations.

**Materials and Methods:** Serial cross-sectional analysis of five waves of the American Board of Family Medicine's Continuous Certification Questionnaire (2022–26). Adjusted logistic regression compared odds of being very satisfied across 11 EHRs; Cochran-Armitage tests assessed vendor trends. Ease of using outside information was analyzed separately for same- and different-vendor sources.

**Results:** Across 36,785 FPs, adjusted odds of being very satisfied relative to Epic ranged from 0.91 (95% CI, 0.74–1.11) for Elation Health to 0.17 (95% CI, 0.15–0.20) for Oracle Health/Cerner. Only Epic's users became significantly more satisfied (Benjamini-Hochberg adjusted $P < .001$), widening the vendor satisfaction gap. Within comparable periods, ease of using different-vendor information did not improve; fewer than 13% rated it very easy. Epic's interoperability advantage reversed by exchange type: non-Epic users had less than half the odds of Epic users of rating same-vendor interoperability very easy (adjusted OR, 0.41; 95% CI, 0.38–0.44) but 1.54 times the odds for different-vendor interoperability (95% CI, 1.35–1.75).

**Discussion:** Rising national exchange volume has not yielded easier use of outside information. Epic's advantage is confined to its own network, so market concentration may shift, rather than solve, the different-vendor problem.

**Conclusion:** The ABFM data are already informing EHR policy but also have utility for the EHR marketplace. While EHR satisfaction has known implications for burden and burnout, poor interoperability is a quality and safety threat that may be compounded by artificial intelligence.

## Background and Significance

Electronic health records (EHRs) are now a constant presence in medicine and primary care physicians carry a heavy documentation burden.[1] How well an EHR works is not a property of the software alone; it varies by vendor, by the setting, and by who is using it.[2,3]

This variation matters for two audiences. The first is policymakers. The 21st Century Cures Act's provisions against information-blocking and requirements for application programming interfaces (APIs) have come fully into effect. The Trusted Exchange Framework and Common Agreement (TEFCA) was published in 2022,[4] and the first Qualified Health Information Networks were designated in late 2023.[5] These policies have substantially expanded the infrastructure and volume of health information exchange, but infrastructure-level statistics on the volume of exchange do not mean that clinicians can find and use that information at the point of care. The value of an EHR also depends on network effects: hospitals are markedly more likely to exchange outside their own system when the surrounding region shares their vendor, and state programs have not reliably overcome that pattern.[6,7] In 2022, only 8% of family physicians found it very easy to access information from an organization using a different EHR, compared with 38% for the same EHR.[8] It is unclear whether TEFCA has improved this situation.

The second audience is the health system or clinic deciding which EHR to use. Choosing an EHR is among the largest and least reversible capital decisions a delivery organization makes. The Veterans Health Administration's replacement of CPRS/VistA illustrates the dangers of changing EHRs: after the 2020 launch, clinicians reported treatment delays, medication errors, and repeated outages, and by 2023 the department had linked the rollout to thousands of documented patient harms.[9] Purchasers have little vendor-specific evidence on clinician experience on which to base those decisions. Annual census-based, vendor-named reporting could change that.

The American Board of Family Medicine's Continuous Certification Questionnaire (CCQ) is an annual survey with a 100% response rate among family physicians who register to continue board certification. In this study, we evaluate changes in EHR experience across five annual cohorts from 2022 to 2026.[8,10] We address three questions. First, how large are the differences in overall EHR satisfaction across vendors, and are they stable? Second, is reported ease of using outside information improving in the wake of TEFCA, and does it differ by vendor? Third, do practice setting and rurality shape satisfaction and reported ease?

## Materials and Methods

### Study design and data source

We analyzed CCQ responses from family physicians in five annual waves, 2022 through 2026, restricted to those reporting direct patient care and outpatient continuity care in the United States. Physicians in each recertification cohort were on a 7- or 10-year cycle; therefore, each wave drew a distinct cohort and the design is a serial cross-section rather than a panel.

### Outcome measures

Our primary outcome throughout was the proportion of respondents reporting the top response category: very satisfied for overall EHR satisfaction, and very easy for each ease item. We chose the top category over a top-two collapse because it was less affected by the 2024 questionnaire revision described below and because top-two already exceeds 80% to 90%, leaving too little room to measure improvement. Top-two results are reported as sensitivity analyses (Supplementary Table S1, Supplementary Figure S1).

Overall EHR satisfaction was measured with a single item on a 5-point Likert scale (very dissatisfied, somewhat dissatisfied, neither satisfied nor dissatisfied, somewhat satisfied, very satisfied). Ease of using outside information was measured with two items on a 3-point scale (not at all easy, somewhat easy, very easy); "not applicable" and "don't know" were excluded from the analytic denominators. From 2024 onward, each ease item was administered only to respondents who reported the ability to exchange medical data (Supplementary Figure S2).

### Vendor and practice characteristics

Respondents selected their EHR from a structured drop-down list. Free-text entries under "Other" were standardized to named vendors or to three residual categories (other named vendor, proprietary or in-house, unknown). Vendors with at least 250 total valid satisfaction responses were modeled individually, yielding 11 named vendors; residual categories are described but not modeled (Supplementary Figure S3).

Rurality was classified from the Rural-Urban Commuting Area code of each respondent's practice ZIP code and dichotomized defining rural as RUCA > 3. Practice setting was classified into five categories: hospital or health system, independent, academic, governmental (which includes federally qualified health centers and look-alikes), and other.

### Statistical analysis

**Vendor differences (question 1: how large are vendor differences in satisfaction?).** We tested the association between vendor and overall satisfaction within each year by chi-square (Fisher's exact where expected counts fell below 5), Kruskal-Wallis tests, and pairwise vendor contrasts with Benjamini-Hochberg (BH) correction. Adjusted differences were estimated with logistic regression of very satisfied on vendor (reference: Epic), rurality, practice setting (reference: hospital or health system), ownership (reference: no ownership stake, 100% employed), practice size, census region, and survey year as a categorical term, using complete cases (n = 29,337). Proportions were computed within each stratum and survey year using 95% Wilson score intervals. To compare predictors without favoring variables with more categories, we computed the change in McFadden pseudo-$R^2$ and in the area under the receiver operating characteristic curve associated with removing each block. To assess sensitivity to the expanded adjustment set, the same model was refitted with setting and rurality only, on the same sample.

**Vendor trajectories (question 1, continued: are those differences stable over time?).** Each eligible vendor's own trend was tested with the Cochran-Armitage test, with BH correction across vendors. Whether trajectories differed was tested with a vendor-by-year interaction and a block likelihood-ratio test, the interaction terms expressing each vendor's per-year slope relative to Epic's. This model was fitted with and without the adjustment covariates. Trend tests additionally required at least 50 respondents every year.

**Interoperability trends and vendor differences (question 2: is reported ease of using outside information changing, and does it differ by vendor?).** Each ease series was analyzed on the denominator of respondents answering that item substantively. Because the 2024 questionnaire revision restricted the ease items to physicians who first reported the ability to exchange, trends were tested separately for 2022–2023 and 2024–2026 using Cochran-Armitage tests, with BH correction across the four series-period combinations. The same tests were run for each vendor with at least 30 such responses in every year of a period. To compare vendors, a logistic model predicted a very easy rating (outcome) from vendor group (Epic versus the ten other modeled vendors, grouped because vendor-year cells were small), practice setting, rurality, and measurement period (predictors); adjusting for period absorbs the level shift from the 2024 questionnaire revision.

**Practice characteristics (question 3: do practice setting and rurality shape satisfaction and reported ease?).** For satisfaction, setting-by-year and rurality-by-year interactions tested whether trends differed by practice type, and vendor-group-by-setting, vendor-group-by-rurality, and rurality-by-setting interactions tested whether practice type modified vendor differences. For interoperability, vendor-group-by-setting and vendor-group-by-rurality interactions were added to the question 2 model with measurement periods pooled (Supplementary Table S2). Interactions were tested by likelihood ratio; where one was significant, the vendor contrast was re-estimated within each stratum.

All tests were two-sided, with significance defined as $P < .05$ after BH correction where applicable. A proportional-odds model over the full five-point satisfaction scale, with the same covariates and a linear year term, was fitted as a sensitivity analysis. In addition, interoperability interactions were tested with periods pooled and within each period (Supplementary Table S2). Analyses used R version 4.5.3, with MASS, dplyr, tidyr, and broom for data handling and model output, and ggplot2 for figures.

**Use of generative artificial intelligence.** Generative AI (Claude, Anthropic) produced draft code for the statistical analysis, tables, and figures and assisted with editing manuscript text. The authors directed all analyses, ran and verified all code on the restricted ABFM data, interpreted the results, reviewed the manuscript, and take full responsibility for its content.

**TABLE 1. Characteristics of Respondents by Survey Year**

| Characteristic | 2022 | 2023 | 2024 | 2025 | 2026 |
|---|---|---|---|---|---|
| N respondents | 4,368 | 8,590 | 8,572 | 7,591 | 7,664 |
| **Age group** | | | | | |
| <40 | 615 (14.1%) | 652 (7.6%) | 761 (8.9%) | 744 (9.8%) | 735 (9.6%) |
| 40-49 | 1908 (43.7%) | 2552 (29.7%) | 2586 (30.2%) | 2398 (31.6%) | 2350 (30.7%) |
| 50-59 | 1080 (24.7%) | 3008 (35.0%) | 2851 (33.3%) | 2548 (33.6%) | 2620 (34.2%) |
| 60+ | 765 (17.5%) | 2378 (27.7%) | 2374 (27.7%) | 1901 (25.0%) | 1956 (25.5%) |
| **Gender** | | | | | |
| Female | 2123 (48.6%) | 3835 (44.6%) | 3914 (45.7%) | 3692 (48.6%) | 3741 (48.8%) |
| Male | 2232 (51.1%) | 4597 (53.5%) | 4596 (53.6%) | 3872 (51.0%) | 3898 (50.9%) |
| Other/Declined | 13 (0.3%) | 158 (1.8%) | 62 (0.7%) | 27 (0.4%) | 25 (0.3%) |
| **Race/ethnicity** | | | | | |
| White | - | 5320 (61.9%) | 5019 (58.6%) | 4681 (61.7%) | 4621 (60.3%) |
| Asian | - | 1400 (16.3%) | 1435 (16.7%) | 1374 (18.1%) | 1417 (18.5%) |
| Hispanic (any race) | - | 672 (7.8%) | 705 (8.2%) | 592 (7.8%) | 581 (7.6%) |
| Black | - | 561 (6.5%) | 609 (7.1%) | 542 (7.1%) | 536 (7.0%) |
| Multiracial | - | 163 (1.9%) | 195 (2.3%) | 65 (0.9%) | 69 (0.9%) |
| MENA | - | 141 (1.6%) | 113 (1.3%) | 44 (0.6%) | 47 (0.6%) |
| AI/AN | - | 46 (0.5%) | 33 (0.4%) | 36 (0.5%) | 36 (0.5%) |
| NH/PI | - | 0 (0.0%) | 28 (0.3%) | 32 (0.4%) | 25 (0.3%) |
| Other | - | 287 (3.3%) | 255 (3.0%) | 92 (1.2%) | 177 (2.3%) |
| Prefer not to answer | - | 0 (0.0%) | 180 (2.1%) | 133 (1.8%) | 149 (1.9%) |
| First-generation college | - | 2436 (28.4%) | 2498 (29.1%) | 2141 (28.2%) | 2135 (27.9%) |
| **Medical background** | | | | | |
| DO, Non-IMG | 584 (13.5%) | 994 (11.7%) | 1035 (12.3%) | 1060 (14.2%) | 1083 (14.4%) |
| MD, IMG | 1359 (31.4%) | 1974 (23.3%) | 2036 (24.2%) | 1755 (23.6%) | 1847 (24.6%) |
| MD, Non-IMG | 2385 (55.1%) | 5512 (65.0%) | 5336 (63.5%) | 4631 (62.2%) | 4588 (61.0%) |
| Years in practice, median (IQR) | 12.0 (10.0, 23.0) | 20.0 (12.0, 27.0) | 20.0 (10.0, 27.0) | 19.0 (10.0, 26.0) | 20.0 (10.0, 26.0) |
| **Faculty appointment** | | | | | |
| No | 3053 (69.9%) | 5905 (68.7%) | 5921 (69.1%) | 5337 (70.3%) | 5396 (70.4%) |
| Yes, clinical faculty | 814 (18.6%) | 1741 (20.3%) | 1713 (20.0%) | 1412 (18.6%) | 1418 (18.5%) |
| Yes, core faculty | 501 (11.5%) | 944 (11.0%) | 938 (10.9%) | 842 (11.1%) | 850 (11.1%) |
| **Census region** | | | | | |
| Midwest | 1037 (23.8%) | 2097 (24.5%) | 1987 (23.2%) | 1817 (24.0%) | 1791 (23.4%) |
| Northeast | 603 (13.8%) | 1158 (13.5%) | 1151 (13.5%) | 1010 (13.3%) | 1058 (13.8%) |

| Characteristic | 2022 | 2023 | 2024 | 2025 | 2026 |
|---|---|---|---|---|---|
| South | 1547 (35.5%) | 3055 (35.7%) | 3068 (35.9%) | 2698 (35.6%) | 2716 (35.5%) |
| West | 1167 (26.8%) | 2252 (26.3%) | 2341 (27.4%) | 2048 (27.0%) | 2080 (27.2%) |
| **Rurality** | | | | | |
| Urban (metro) | 3679 (84.2%) | 7060 (82.2%) | 7080 (82.6%) | 6358 (83.8%) | 6432 (83.9%) |
| Rural (non-metro) | 689 (15.8%) | 1530 (17.8%) | 1492 (17.4%) | 1233 (16.2%) | 1232 (16.1%) |
| **Practices in a Health Professional Shortage Area** | 1705 (39.0%) | 3847 (44.8%) | 3859 (45.0%) | 3214 (42.3%) | 3246 (42.4%) |
| **Practice site type** | | | | | |
| Hospital/health system | 1514 (34.7%) | 3045 (35.4%) | 2931 (34.2%) | 2450 (32.3%) | 2482 (32.4%) |
| Independent | 1230 (28.2%) | 2628 (30.6%) | 2547 (29.7%) | 2098 (27.6%) | 2137 (27.9%) |
| Academic | 317 (7.3%) | 601 (7.0%) | 614 (7.2%) | 576 (7.6%) | 589 (7.7%) |
| Governmental | 712 (16.3%) | 1312 (15.3%) | 1398 (16.3%) | 1234 (16.3%) | 1230 (16.0%) |
| Other | 595 (13.6%) | 1004 (11.7%) | 1082 (12.6%) | 1233 (16.2%) | 1226 (16.0%) |
| **Site ownership** | | | | | |
| No official ownership stake | 2985 (68.3%) | 5745 (66.9%) | 5789 (67.5%) | 4234 (64.3%) | 4249 (64.2%) |
| Partial owner or shareholder | 644 (14.7%) | 1359 (15.8%) | 1235 (14.4%) | 1075 (16.3%) | 1064 (16.1%) |
| Sole owner | 512 (11.7%) | 1109 (12.9%) | 1133 (13.2%) | 947 (14.4%) | 976 (14.7%) |
| Self-employed as a contractor | 160 (3.7%) | 261 (3.0%) | 280 (3.3%) | 212 (3.2%) | 206 (3.1%) |
| Other | 67 (1.5%) | 116 (1.4%) | 135 (1.6%) | 112 (1.7%) | 123 (1.9%) |
| **Site size** | | | | | |
| Solo practice | 452 (10.3%) | 978 (11.4%) | 962 (11.2%) | 832 (11.0%) | 833 (10.9%) |
| 2-5 clinicians | 1429 (32.7%) | 2869 (33.4%) | 2757 (32.2%) | 2466 (32.5%) | 2436 (31.8%) |
| 6-20 clinicians | 1337 (30.6%) | 2697 (31.4%) | 2730 (31.8%) | 2492 (32.8%) | 2578 (33.6%) |
| >20 clinicians | 1150 (26.3%) | 2046 (23.8%) | 2123 (24.8%) | 1801 (23.7%) | 1817 (23.7%) |

Values are n (%) unless noted; percentages exclude missing responses. Race/ethnicity and first-generation college status were not collected in 2022. IMG, international medical graduate; MENA, Middle Eastern or North African; AI/AN, American Indian or Alaska Native; NH/PI, Native Hawaiian or Pacific Islander; IQR, interquartile range.

## Results

### Respondent characteristics

Across the five waves, 36,785 family physicians met inclusion criteria: 4,368 in 2022, 8,590 in 2023, 8,572 in 2024, 7,591 in 2025, and 7,664 in 2026 (Table 1). Demographic and practice characteristics were largely stable year to year, with one exception: 2022 respondents were younger and less experienced, reflecting a change to the examination schedule rather than a shift in the workforce. Epic was the modal EHR, rising from 36% of respondents in 2022 to 46% in 2026, followed by eClinicalWorks, AthenaHealth, Oracle Health/Cerner, and Veradigm (Allscripts).

### Differences in overall satisfaction between vendors

Which EHR a physician used mattered far more for satisfaction than any practice characteristic we measured. Removing the vendor term from the adjusted model reduced explanatory power more than 20 times as much as removing ownership, the next most informative variable, and roughly 45 times as much as removing practice setting (Supplementary Table S3). Satisfaction differed across vendors in every survey year (all $P < .001$).

Only one vendor was statistically similar to Epic, and the rest trailed it substantially. In the adjusted model, the odds of being very satisfied relative to Epic were 0.91 (95% CI, 0.74 to 1.11) for Elation Health and 0.64 (95% CI, 0.59 to 0.70) for AthenaHealth. The remaining eight fell in a broad low band running from 0.39 for Veterans Affairs CPRS/VistA down to 0.17 for Oracle Health/Cerner and MEDITECH, all $P < .001$ (Figure 1A, Supplementary Table S4). Year-by-year percentages for every vendor are in Table 2.

**TABLE 2. Very Satisfied With the EHR, by Vendor and Survey Year**

| Vendor | Total N | 2022 | 2023 | 2024 | 2025 | 2026 |
|---|---|---|---|---|---|---|
| Epic | 15,237 | 36.6 (34.2–39.0) | 33.9 (32.3–35.6) | 35.4 (33.8–37.0) | 41.8 (40.1–43.4) | 43.7 (42.1–45.4) |
| Elation Health | 437 | 38.9 (24.8–55.1) | 44.2 (34.6–54.2) | 40.4 (31.5–50.0) | 52.0 (42.3–61.7) | 48.1 (38.7–57.6) |
| AthenaHealth | 3,528 | 31.7 (27.5–36.2) | 30.4 (27.4–33.6) | 32.8 (29.7–36.1) | 30.3 (27.0–33.8) | 33.7 (30.3–37.2) |
| Veterans Affairs (CPRS/VistA) | 806 | 22.4 (15.6–31.2) | 18.1 (13.1–24.4) | 15.0 (11.0–20.3) | 18.1 (12.8–25.1) | 19.0 (13.5–26.2) |
| eClinicalWorks | 3,687 | 14.7 (11.8–18.2) | 17.2 (14.9–19.7) | 18.6 (16.1–21.3) | 18.3 (15.6–21.3) | 18.0 (15.4–21.0) |
| Veradigm (Allscripts) | 1,970 | 16.4 (12.7–20.9) | 15.0 (12.2–18.2) | 17.9 (14.7–21.6) | 15.3 (11.9–19.6) | 17.3 (13.3–22.1) |
| CompuGroup (CGM) | 450 | 13.2 (6.5–24.8) | 18.3 (12.6–25.8) | 19.8 (13.2–28.6) | 22.5 (14.7–32.8) | 14.1 (8.3–23.1) |
| NextGen | 1,324 | 17.7 (12.9–23.9) | 11.3 (8.4–15.1) | 10.1 (7.3–13.7) | 7.9 (5.2–11.9) | 10.3 (6.8–15.4) |
| MEDITECH | 511 | 11.5 (5.4–23.0) | 9.3 (5.0–16.7) | 7.9 (4.4–14.0) | 7.7 (4.3–13.2) | 15.1 (9.2–23.7) |
| Greenway Health | 471 | 9.8 (5.0–18.1) | 12.5 (8.0–19.1) | 10.1 (5.7–17.2) | 11.5 (6.2–20.5) | 12.1 (6.3–22.1) |
| Oracle Health / Cerner | 2,718 | 10.1 (7.3–13.7) | 9.8 (7.8–12.4) | 8.4 (6.5–10.8) | 7.8 (5.9–10.3) | 9.2 (7.0–11.9) |
| **Residual categories (descriptive only)** | | | | | | |
| Other named vendor | 3,122 | 21.3 (17.4–25.7) | 21.6 (18.7–24.8) | 22.8 (19.8–26.1) | 22.0 (19.0–25.3) | 23.8 (20.7–27.2) |
| Proprietary/ In-house | 77 | – | 13.8 (5.5–30.6) | 17.4 (7.0–37.1) | – | – |
| Unknown/ Unclear | 664 | 20.0 (13.0–29.4) | 18.2 (12.3–26.0) | 20.4 (14.6–27.8) | 17.7 (12.4–24.7) | 20.1 (14.7–26.9) |

Values are percentage very satisfied (95% Wilson CI) among respondents answering the item. –, suppressed (n < 15). Residual categories were not modeled.

Greenway Health, MEDITECH, and NextGen were statistically indistinguishable from one another (Supplementary Table S4), so the sequence in which they appear in Figure 1A carries no information and should not be read as a ranking. The low-band systems differ from Epic at both ends of the scale: their users more often report dissatisfaction and far less often select very satisfied (Figure 2).

**FIGURE 1: Adjusted Vendor Differences in Satisfaction Level (A) and Trend (B)**

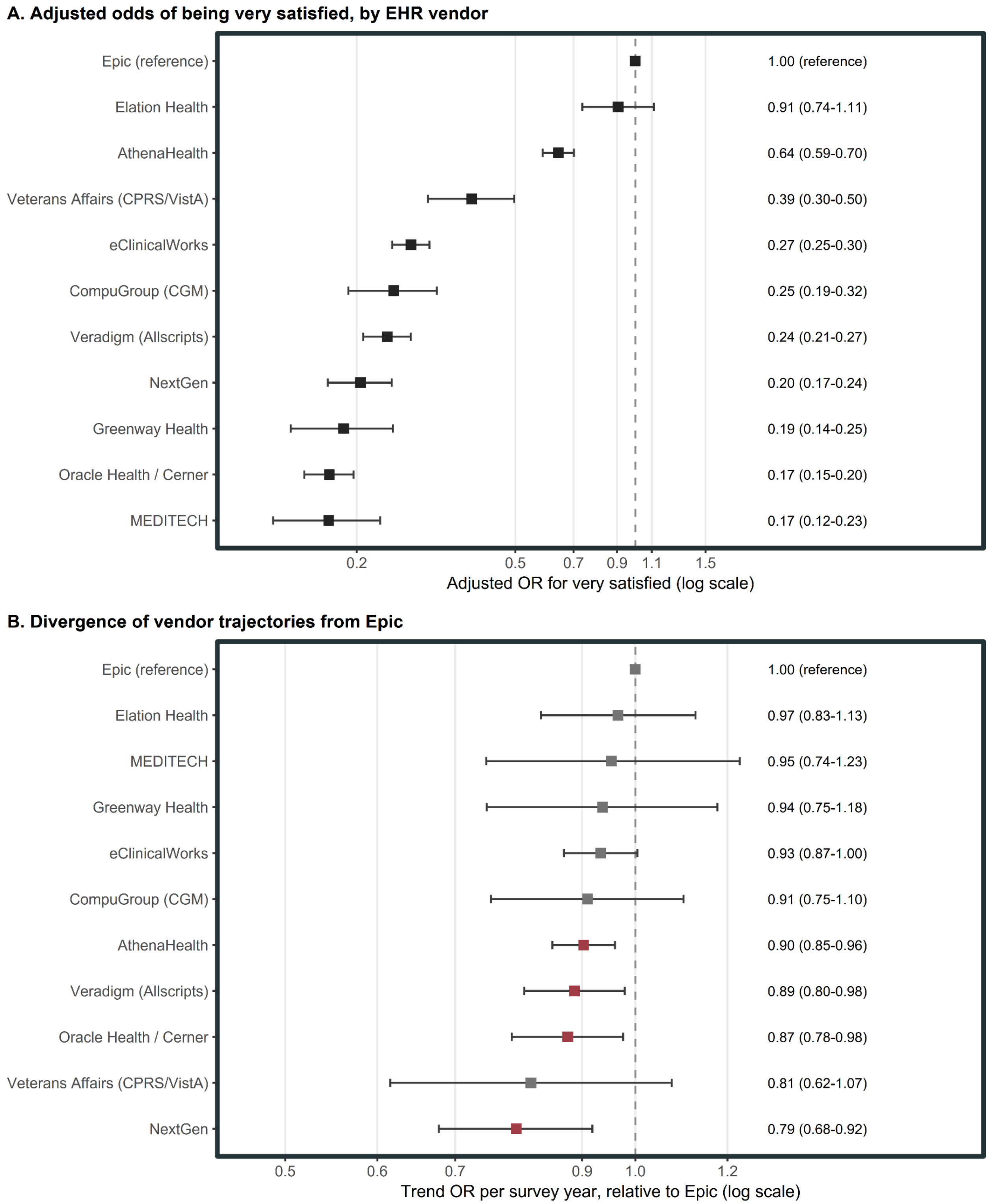

**Figure 1.** Odds ratios (95% CI) for being very satisfied, relative to Epic, adjusted for rurality, practice setting, ownership, practice size, census region, and survey year. A, level; B, ratio of each vendor's per-year trend to Epic's (below 1, falling behind Epic; red, CI excludes 1). Log scale; panels use different axis ranges.

**FIGURE 2: Distribution of EHR Satisfaction Responses by Vendor**

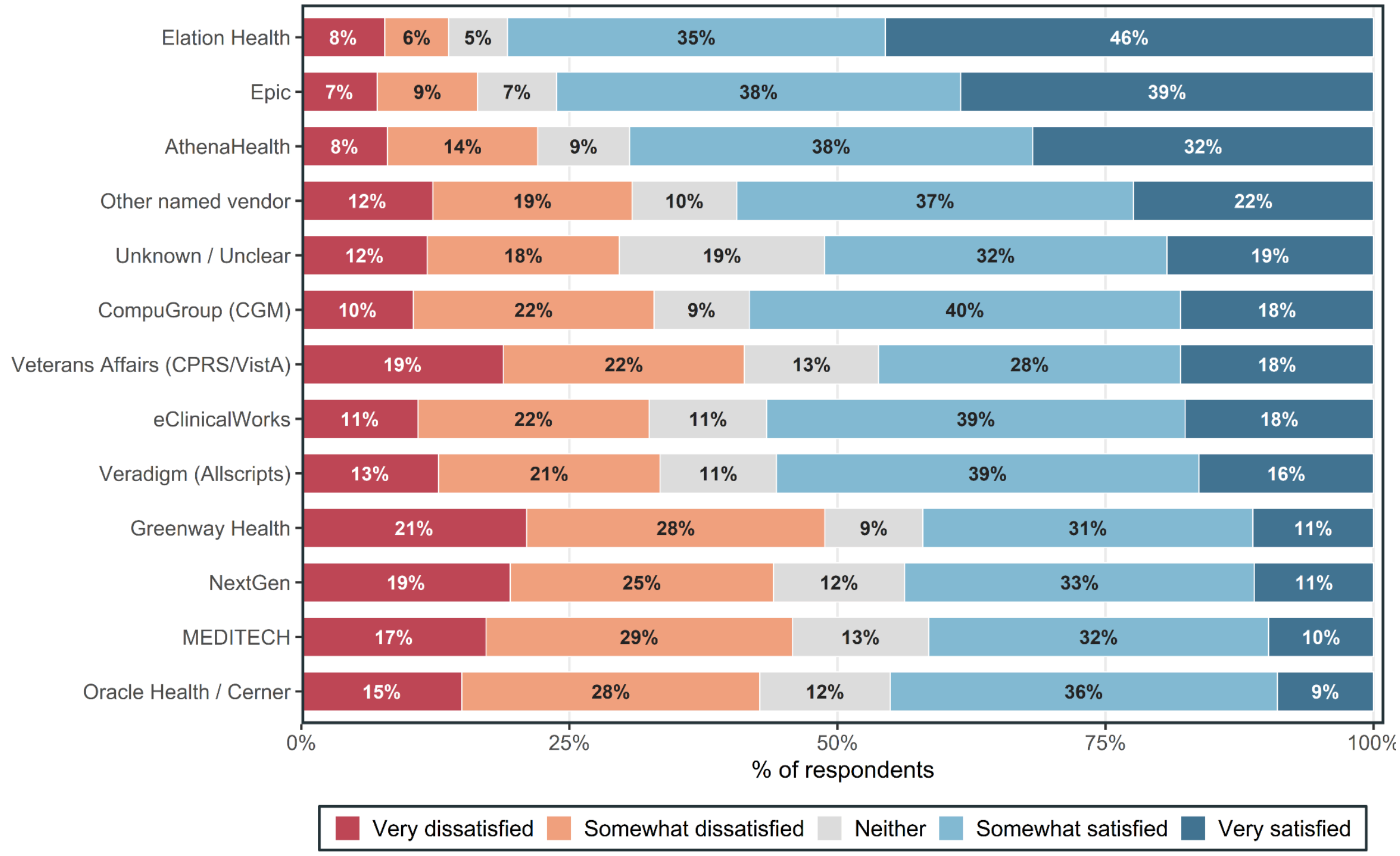

**Figure 2.** Percentage of respondents at each level of the five-point satisfaction item, pooled 2022–2026, ordered by percentage very satisfied. Vendors with fewer than 100 responses are omitted; residual categories are shown for reference.

Vendor differences were also strongly modified by practice setting (interaction P < .001; Table 3). In hospital or health system settings, users of other vendors had roughly one-quarter the odds of Epic users of being very satisfied (OR, 0.27; 95% CI, 0.24 to 0.30), whereas in independent practice the disadvantage was attenuated about 2.7-fold, to 0.72 (95% CI, 0.63 to 0.82). No other setting differed significantly from hospital or health system practice. With ownership, practice size, and region in the model, independent practice carried no satisfaction advantage of its own (OR, 0.98; 95% CI, 0.89 to 1.09, against 1.33 with setting and rurality alone; Supplementary Table S4). Sole owners (OR, 1.92; 95% CI, 1.67 to 2.21) and partial owners (1.25; 1.14 to 1.37) had higher odds than employed physicians of being very satisfied (Supplementary Table S4). Rurality showed no association after adjustment (OR, 1.00; 95% CI, 0.92 to 1.08; Supplementary Table S4).

**TABLE 3. Exploratory Interactions in the Odds of Being Very Satisfied, 2022–2026**

| Term | OR (95% CI) | P value |
|---|---|---|
| **Epic vs other vendors, by practice setting (block LRT, BH-adjusted P <.001*)** | | |
| Other vendors (vs Epic) | 0.27 (0.24–0.30) | <.001* |
| Other vendors × independent practice | 2.65 (2.24–3.15) | <.001* |
| Other vendors × academic practice | 0.78 (0.59–1.03) | .08 |
| Other vendors × governmental practice | 1.02 (0.85–1.22) | .82 |
| Other vendors × other practice setting | 0.97 (0.81–1.16) | .75 |
| **Rural vs urban, by practice setting (block LRT, BH-adjusted P = .002*)** | | |
| Rural (vs urban) | 0.90 (0.81–1.00) | .05 |
| Rural × independent practice | 1.48 (1.23–1.78) | <.001* |
| Rural × academic practice | 1.03 (0.66–1.58) | .90 |
| Rural × governmental practice | 1.06 (0.87–1.28) | .56 |
| Rural × other practice setting | 0.97 (0.68–1.36) | .86 |
| Other vendors × rural practice (single-term test) | 1.04 (0.90–1.20) | .60 |

Logistic models of very satisfied, adjusted for survey year (linear) and the remaining covariate. References: Epic and hospital or health system practice. Each block's first row is the effect in hospital or health system practice; effects in other settings are the product of the two relevant rows. OR, odds ratio; CI, confidence interval. Post hoc and hypothesis-generating. LRT, likelihood-ratio test; BH, Benjamini-Hochberg. *P < .05.

## Vendor trajectories, 2022–2026

Only Epic's users consistently became more satisfied over the five years. Odds of a very satisfied response for physicians using Epic rose by 12% per year (OR, 1.12; 95% CI, 1.09 to 1.15; BH-adjusted P < .001; Supplementary Table S1), and no other vendor's trend survived correction. Users of Elation Health, excluded from trend testing by the sample-size rule, reported between 39% and 52% very satisfied without clear direction (Figure 3).

**FIGURE 3: Trends in Satisfaction with EHR by Vendors, 2022–2026**

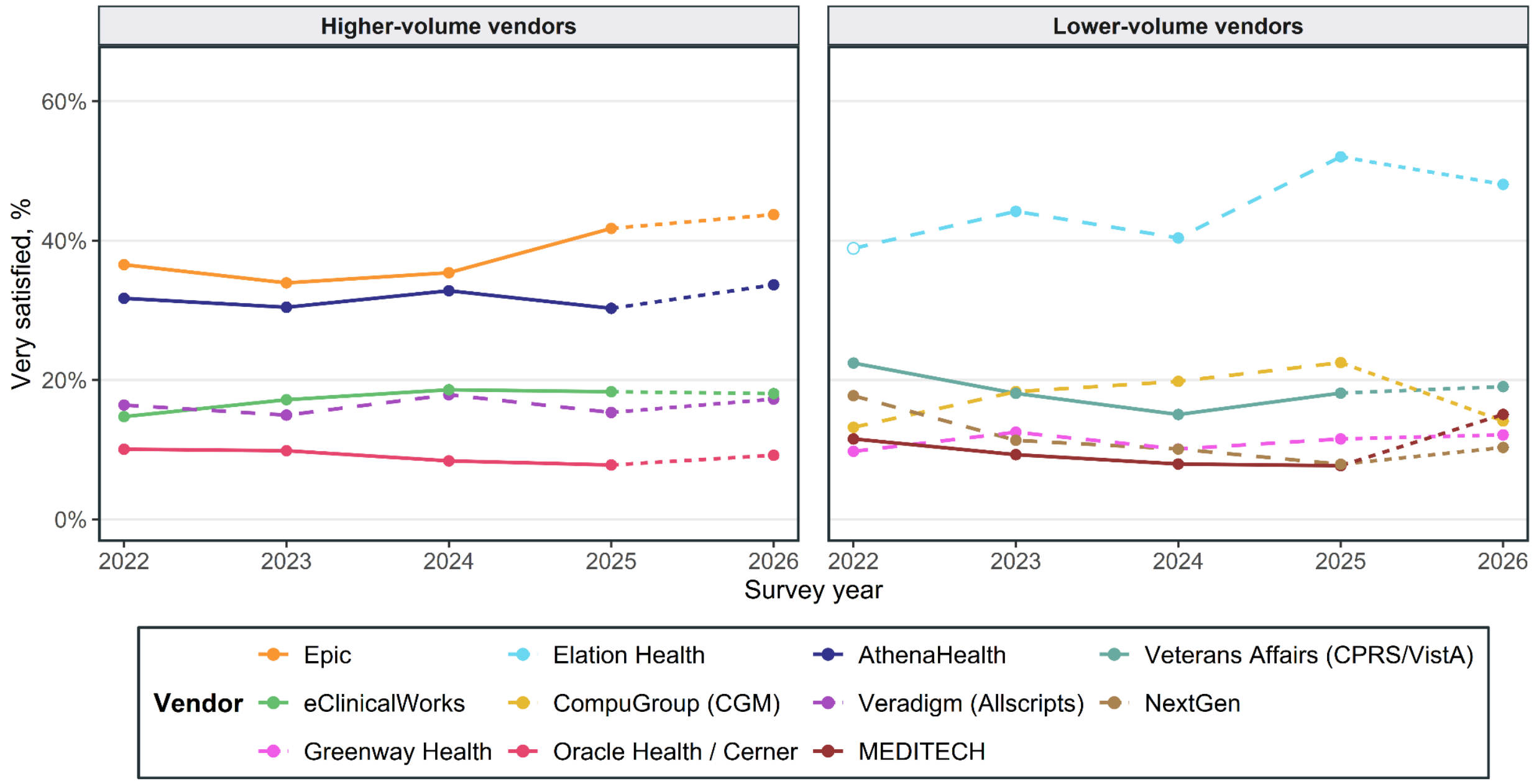

**Figure 3.** Percentage very satisfied by vendor and survey year. Hollow points, n = 15–49. Dashed segment, 2026 wave.

Because only Epic improved, the differences between vendors widened (P < .001 for the vendor-by-year interaction). In the adjusted model, four systems fell measurably behind Epic's pace: NextGen (trend OR, 0.79 per year relative to Epic), Oracle Health/Cerner (0.87), Veradigm (0.89), and AthenaHealth (0.90) (Figure 1B, Supplementary Table S1). This pattern is not explained by a changing Epic user base: adjustment left Epic's own trend unchanged (OR, 1.12 per year) and moved every divergence estimate by 0.06 or less, although Veterans Affairs CPRS/VistA, borderline significant unadjusted (0.87), was not after adjustment (0.81; 95% CI, 0.62 to 1.07).

## Interoperability

Reported ease of using outside information did not improve within either measurement period. Different-vendor information was rated very easy by 8.8% and 7.2% of respondents in 2022 and 2023, and by 12.6%, 11.5%, and 12.2% from 2024 through 2026. Same-vendor figures were 42.3% and 45.3%, then 40.3%, 41.0%, and 41.2% (Figure 4). No within-period trend was significant after correction at either ease threshold (Supplementary Table S5), and among the seven to nine vendors eligible for testing in each period only one trend was significant: same-vendor ease declined among Oracle Health/Cerner users between 2024 and 2026, from 32.7% to 21.5% (BH-adjusted P = .02; Supplementary Table S6; Supplementary Figures S4 and S5). The higher different-vendor level from 2024 coincided with the questionnaire revision and should not be read as improvement (Supplementary Table S7).

**FIGURE 4: Trends in Ease of Using Same- and Different-Vendor Information, 2022–2026**

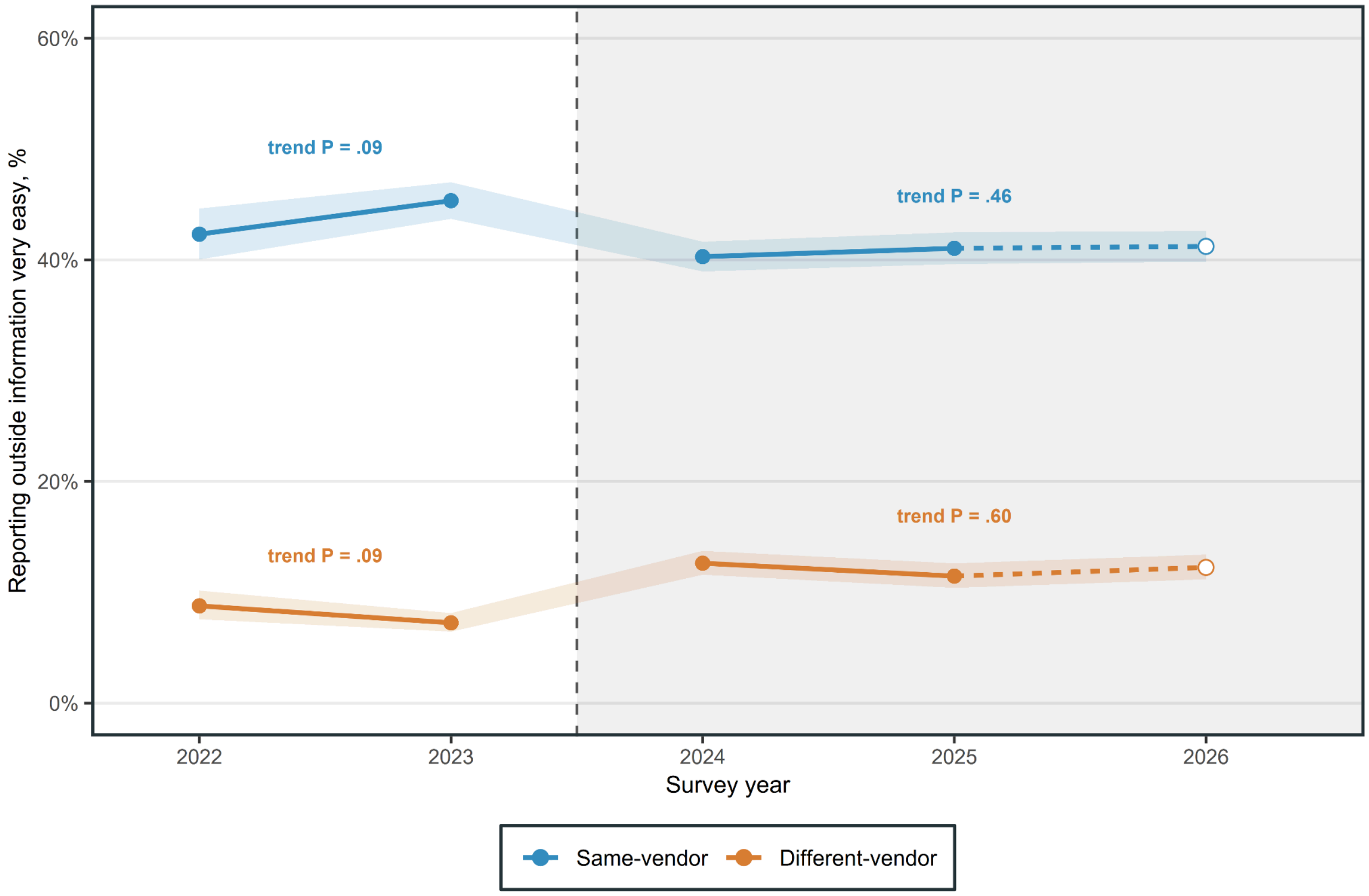

**Figure 4.** Percentage rating outside information very easy, by exchange type. Bands, 95% Wilson CI. The shaded region follows the 2024 questionnaire revision, after which only respondents reporting the ability to exchange were asked; years are compared within periods only. Trend P values are Benjamini-Hochberg adjusted Cochran-Armitage tests. Dashed segment, 2026 wave.

**FIGURE 5: Ease of Using Same- and Different-Vendor Information by Practice Setting and Vendor Group, 2022–2026**

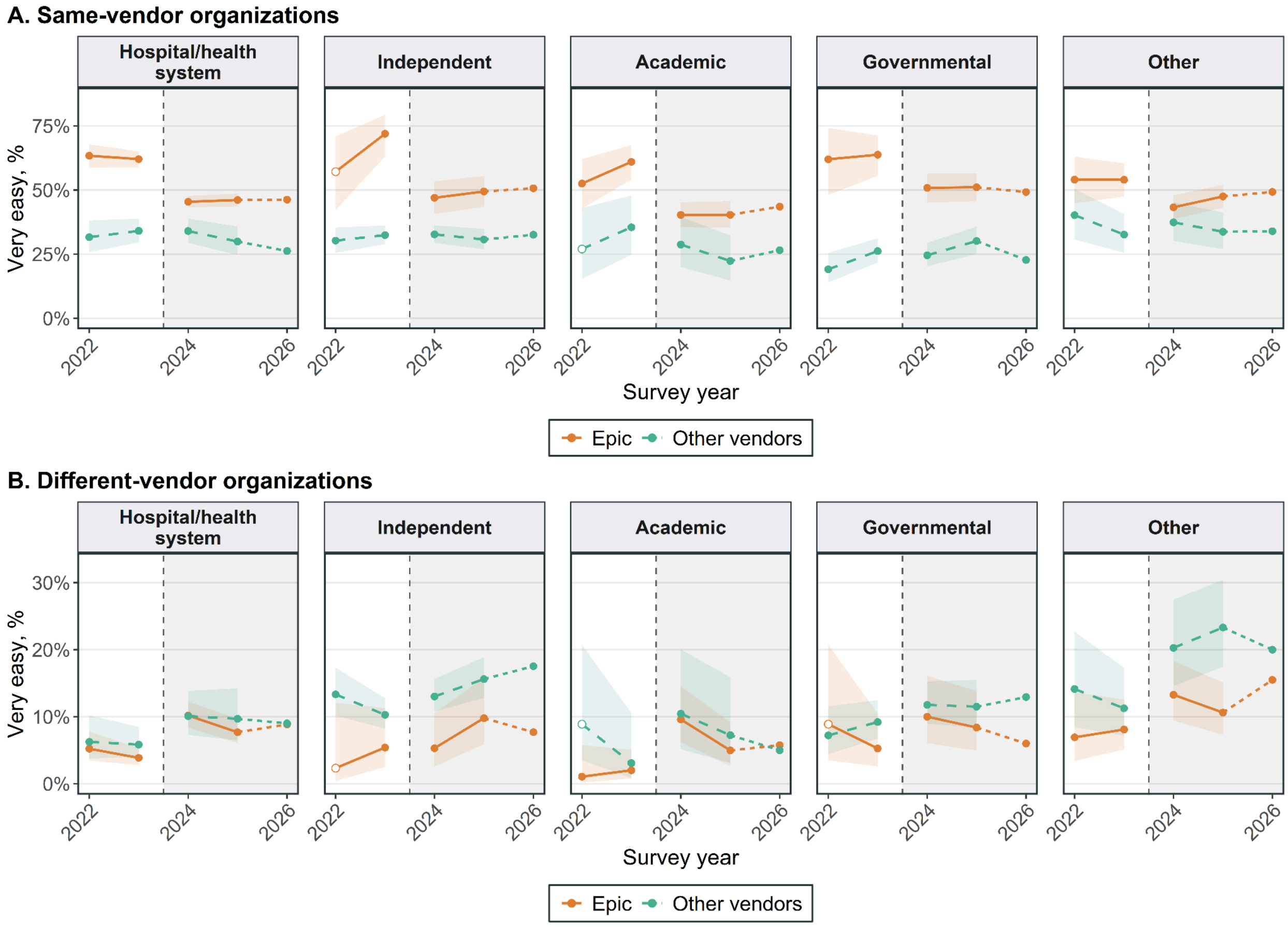

**Figure 5.** Percentage rating outside information very easy, Epic versus other vendors, by practice setting. A, same-vendor; B, different-vendor. Bands, 95% Wilson CI; shading marks the 2024 revision. Panels use separate y-axis ranges. Hollow points, n = 15–49.

Epic's advantage reversed by exchange type (Table 4). Users of other vendors had less than half the odds of Epic users of rating same-vendor information very easy (adjusted OR, 0.41; 95% CI, 0.38 to 0.44), but higher odds of rating different-vendor information very easy (adjusted OR, 1.54; 95% CI, 1.35 to 1.75).

**TABLE 4. Adjusted Odds of Rating Outside Information Very Easy to Use, Other Vendors Versus Epic, 2022–2026**

| Stratum | Same-vendor n (Epic/other) | Same-vendor OR (95% CI) | Different-vendor n (Epic/other) | Different-vendor OR (95% CI) |
|---|---|---|---|---|
| All physicians | 11,757 / 6,789 | 0.41 (0.38–0.44)* | 7,154 / 6,986 | 1.54 (1.35–1.75)* |
| **Practice setting** | | | | |
| Hospital / health system | 6,477 / 1,522 | 0.44 (0.39–0.50)* | 4,004 / 1,428 | 1.17 (0.93–1.47) |
| Independent | 937 / 2,812 | 0.40 (0.34–0.47)* | 598 / 2,926 | 2.29 (1.66–3.25)* |
| Academic | 1,412 / 328 | 0.45 (0.35–0.59)* | 863 / 306 | 1.39 (0.80–2.35) |
| Governmental | 1,125 / 1,411 | 0.28 (0.24–0.33)* | 623 / 1,637 | 1.54 (1.11–2.17)* |
| Other | 1,806 / 716 | 0.55 (0.46–0.66)* | 1,066 / 689 | 1.75 (1.33–2.29)* |
| **Rurality** | | | | |
| Urban (metro) | 10,228 / 5,580 | 0.41 (0.38–0.45)* | 6,199 / 5,648 | 1.50 (1.31–1.72)* |
| Rural (non-metro) | 1,529 / 1,209 | 0.42 (0.35–0.49)* | 955 / 1,338 | 1.77 (1.29–2.46)* |

Odds ratios for users of the ten other modeled vendors relative to Epic, adjusted for practice setting, rurality, and measurement period. Stratum estimates come from models refitted with that stratum as reference. n (Epic/other), respondents using Epic or other vendors; OR, odds ratio; CI, confidence interval. *CI excludes 1.

Practice setting modified both patterns, and rurality modified neither ($P = .90$ and $.34$; Supplementary Table S2; Supplementary Figure S6). The same-vendor deficit relative to Epic was present in every setting, from 0.28 (95% CI, 0.24 to 0.33) in governmental practice to 0.55 (95% CI, 0.46 to 0.66) in other settings (vendor group by setting, BH-adjusted $P < .001$). The different-vendor reversal was not general but concentrated (BH-adjusted $P = .04$): strongest in independent practice (OR, 2.29; 95% CI, 1.66 to 3.25), also present in other (1.75) and governmental settings (1.54), and absent in hospital or health system practice (OR, 1.17; 95% CI, 0.93 to 1.47) and academic practice (1.39; 95% CI, 0.80 to 2.35). Figure 5 shows both contrasts by setting.

## Practice setting, rurality, and satisfaction trajectories

Independent, governmental, and other settings improved more slowly than hospital or health system practice. The setting-by-year interaction was significant ($P = .02$): against a baseline improvement of about 10% per year in hospital and health system settings, independent (trend OR, 0.95 relative to hospital), governmental (0.92), and other settings (0.94) all improved more slowly, and academic practice did not differ (Supplementary Table S8). Rural and urban practices did not differ in their rate of improvement ($P = .15$).

**FIGURE 6: Trends in Satisfaction by Practice Setting, Rurality, and Vendor Group, 2022–2026**

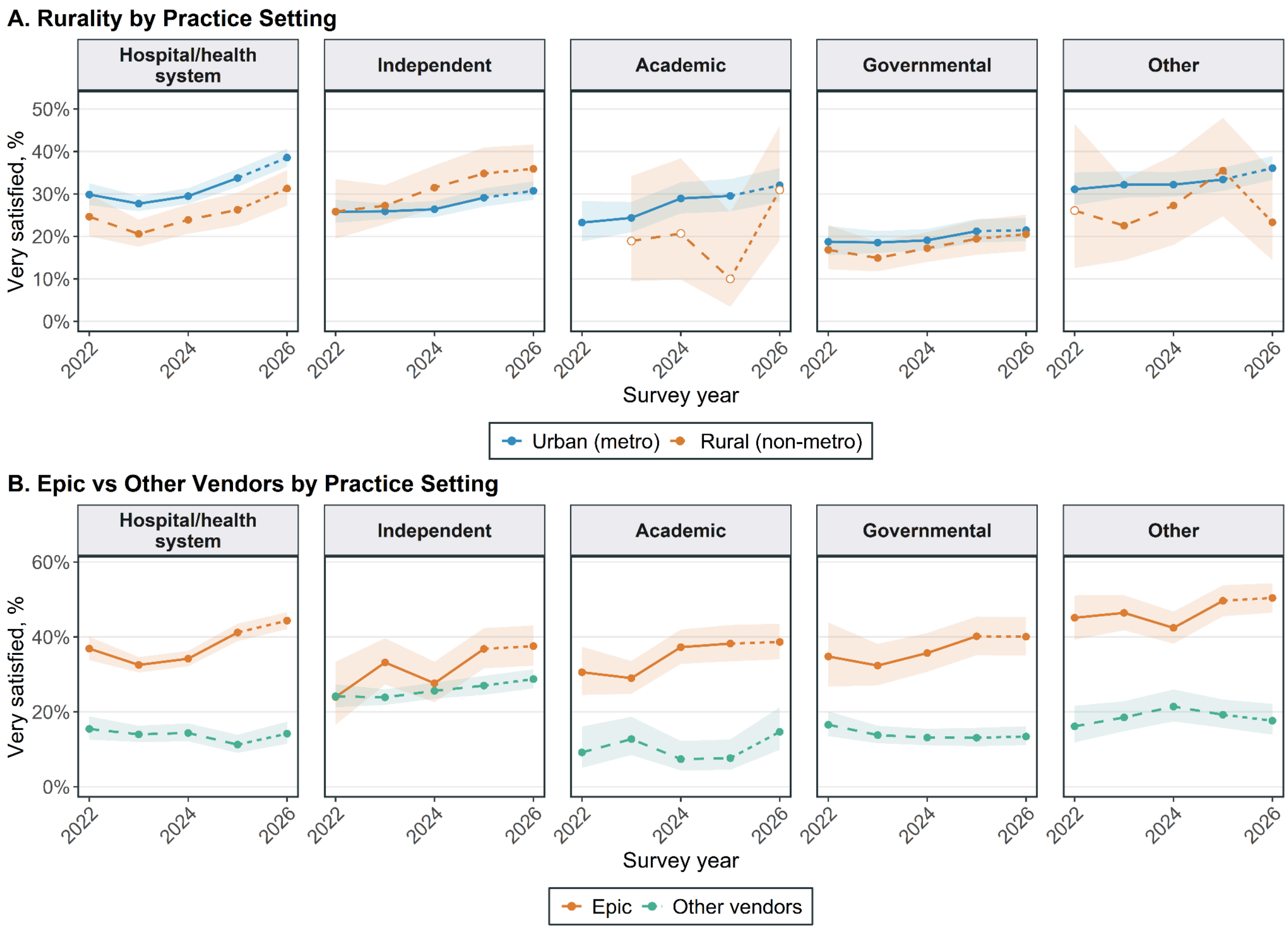

**Figure 6.** Percentage very satisfied by practice setting. A, urban versus rural; B, Epic versus other vendors. Bands, 95% Wilson CI. Hollow points, n = 15–49.

Differences between practice types held steady: neither the rurality-by-setting nor the vendor-group-by-setting difference changed measurably over the study period (P = .74 and .32). Rurality also interacted with setting, again favoring independent practice (BH-adjusted P = .002): the rural–urban odds ratio was 1.48 times higher in independent than in hospital or health system practice (95% CI, 1.23 to 1.78), where rural physicians had slightly lower odds (OR, 0.90; 95% CI, 0.81 to 1.00). Vendor group and rurality did not interact (P = .60). These post hoc analyses are hypothesis-generating (Table 3).

## Sensitivity analyses

All main findings held under sensitivity analyses. A top-two definition of satisfaction preserves the vendor ranking, the widening spread (P < .001), and the vendor-by-setting interaction (P < .001), and additionally shows a significant decline for NextGen (BH-adjusted P < .001) that the primary analysis does not detect. An ordinal model over the full five-point scale identified the same leading systems and the same low band; Oracle Health/Cerner's disadvantage was smaller (OR, 0.26) and Veterans Affairs CPRS/VistA's larger (0.31), because Oracle's gap lies at the top of the scale while VA users more often report dissatisfaction (Figure 2, Supplementary Table S4). The interoperability conclusions hold at either ease threshold (Supplementary Table S5).

Expanding the adjustment set to ownership, practice size, and region moved vendor odds ratios by up to 17%, away from parity with Epic, without materially changing their order (Supplementary Table S4). Within each measurement period, the same-vendor setting interaction held; the different-vendor interaction did not (Supplementary Table S2).

## Discussion

### Principal findings

Across five years, EHR satisfaction was essentially flat for every vendor except Epic, whose users grew steadily more satisfied, so vendor differences widened, with four systems falling measurably behind Epic. Reported ease of using outside information did not improve within either measurement period, and Epic's advantage held for same-vendor exchange but reversed for different-vendor exchange. Practice setting shaped both outcomes, with satisfaction improving more slowly outside hospital or health system practice and the different-vendor reversal strongest in independent practice, while rurality mattered little.

### Who experiences the different-vendor problem

Epic's interoperability advantage is specific to same-vendor exchange. Users of other vendors had less than half the odds of Epic users of finding same-vendor information very easy, and that deficit held in every practice setting and in each measurement period. For different-vendor exchange the association reversed: users of other vendors had about 50% higher odds of rating that information very easy. The effect size is not uniform: it is largest in independent practice and absent in hospital or health system practice, where Epic is most common. One plausible reading is that physicians on Epic meet different-vendor exchange only where their own network does not reach, while those on smaller systems do it routinely; market concentration may therefore change who encounters different-vendor exchange, and when, rather than eliminating it. Because the different-vendor interaction reached significance only with measurement periods pooled, it should be confirmed as the revised questionnaire accumulates waves.

### Why Epic's users became more satisfied

Satisfaction with technology reflects much more than the product itself, including users' expectations, training, and local configuration, none of which this survey measures. We therefore cannot say what about Epic's product or implementation accounts for its users' rising satisfaction; adjusting for setting, rurality, ownership, size, and region did not attenuate the trend. The more striking pattern may be the opposite one: satisfaction with every other vendor stayed essentially flat for five years.

### What this means for a practice choosing a system

The differences in satisfaction and in ease of using outside information, sustained over five years, are a durable feature of the market. Nothing here suggests a low-scoring system will converge on the highest-rated systems within the life of a typical contract.

Comparisons across EHRs should account for the fact that different practice types gravitate toward different systems. Epic's satisfaction advantage is much smaller in independent practice than in hospital or health system practice, so a ranking built mainly on hospital or health system users overstates what an independent practice should expect from Epic. Within the low band the ordering carries no information, and Elation Health's near parity with Epic rests on roughly 100 respondents a year, so it is promising rather than established.

Different-vendor information is hard to use for users of every system, so practices anticipating heavy different-vendor exchange should plan for friction regardless of the system they buy.

Because same-vendor exchange was rated far easier, the vendors used by a practice's referral partners may shape its experience as much as the product it buys; a practice whose partners use a different vendor should not assume Epic's same-vendor advantage will transfer.

### Comparison with prior work

This study adds a longitudinal dimension to a predominantly cross-sectional literature. We have previously reported cross-sectional analyses; here we move to longitudinal analysis in support of monitoring experience amid a changing policy landscape.[8,10] Prior work established the breadth of vendor variation at single points in time;[1,2] we show that it widened over five years. Everson and colleagues documented the same-vendor advantage in 2022,[8] which we extend through 2026. Physicians in physician-owned practices have reported higher EHR satisfaction.[11] If owners rate more favorably in general, part of that difference may reflect response tendency rather than the product.

### Exchange volume versus reported ease

Volume and experience measure different things, and only one has improved. Records exchanged through the national framework grew roughly fiftyfold in about a year, from around 10 million in January 2025 to nearly 500 million by early 2026.[12] Over the same series, the share of physicians finding different-vendor information very easy did not move. Increasing exchange alone does not resolve the last mile challenge of making exchanged information readily usable at the point of care. The gap may become more consequential as AI-enabled clinical decision support comes to depend on longitudinal information assembled across organizations. Growth in volume without matching improvement in physicians' experience is therefore not only a usability problem but a potential constraint on safe and effective AI-enabled care.

### Limitations

No physician was randomly assigned an EHR. Practices choose systems partly by size, capital, ownership, and health-system affiliation, and although we adjusted for setting, rurality, ownership, size, and region, capital, information-technology staffing, local customization, and physicians' input into the choice were unmeasured. The odds ratios therefore describe satisfaction among each system's current users, not the effect of switching systems.

The interoperability analysis carries further constraints. The 2024 questionnaire revision divides the ease series into windows of two and three waves, so an absent trend within either period is insufficient evidence rather than demonstrated stability. Because the ability-to-exchange item was not fielded before 2024, the 2022–2023 waves cannot be restricted the same way, and the vendor-group models pool across a revision that affected vendors unequally in the same-vendor series. Vendor-year cells are also too small to model the 11 vendors individually, hence the Epic-versus-other-vendors contrast, which cannot say whether the reversal belongs to one comparator system or to all ten. Finally, the study describes family physicians only; and the sample comprises those in active outpatient practice who continued certification, so physicians who left are absent, which would understate the worst systems most.

**Conclusion**
Five years after the changes to health information exchange brought about by the Cures Act, satisfaction differences between EHRs are large, durable, and widening, and clinicians' reported ease of using information from other vendors has not improved. Rapid growth in national exchange volume without matching improvement in reported ease suggests that exchange volume alone is an insufficient measure of interoperability success, an issue that will become more consequential as AI tools come to depend on distributed clinical data. Practices make EHR decisions they will live with for a decade, yet typically do so without objective vendor-specific evidence on clinician experience. Trustworthy vendor-named, measurement-stable, annual reporting is feasible and can inform purchasing decisions and policy.

## Statements and Disclosures

**Ethics Statement:** IRB of American Academy of Family Physicians waived ethical approval for this work.

**Funding:** This project is supported by the Office of the National Coordinator for Health Information Technology (ONC) of the U.S. Department of Health and Human Services (HHS) under grant number 90AX0043 for project OPUS-HIT (Outpatient Physician Use and Satisfaction with Health IT). This information or content and conclusions are those of the authors and should not be construed as the official position or policy of, nor should any endorsements be inferred by ONC, HHS, or the U.S. Government.

**Competing Interests:** All authors report no conflicts of interest.

**Data Availability Statement:** Complete survey data are available for projects approved by the American Board of Family Medicine. Please contact the corresponding author to apply for access.

**AI Disclosure:** See Materials and Methods for disclosure of AI use.

## Appendix

### Figure S1:

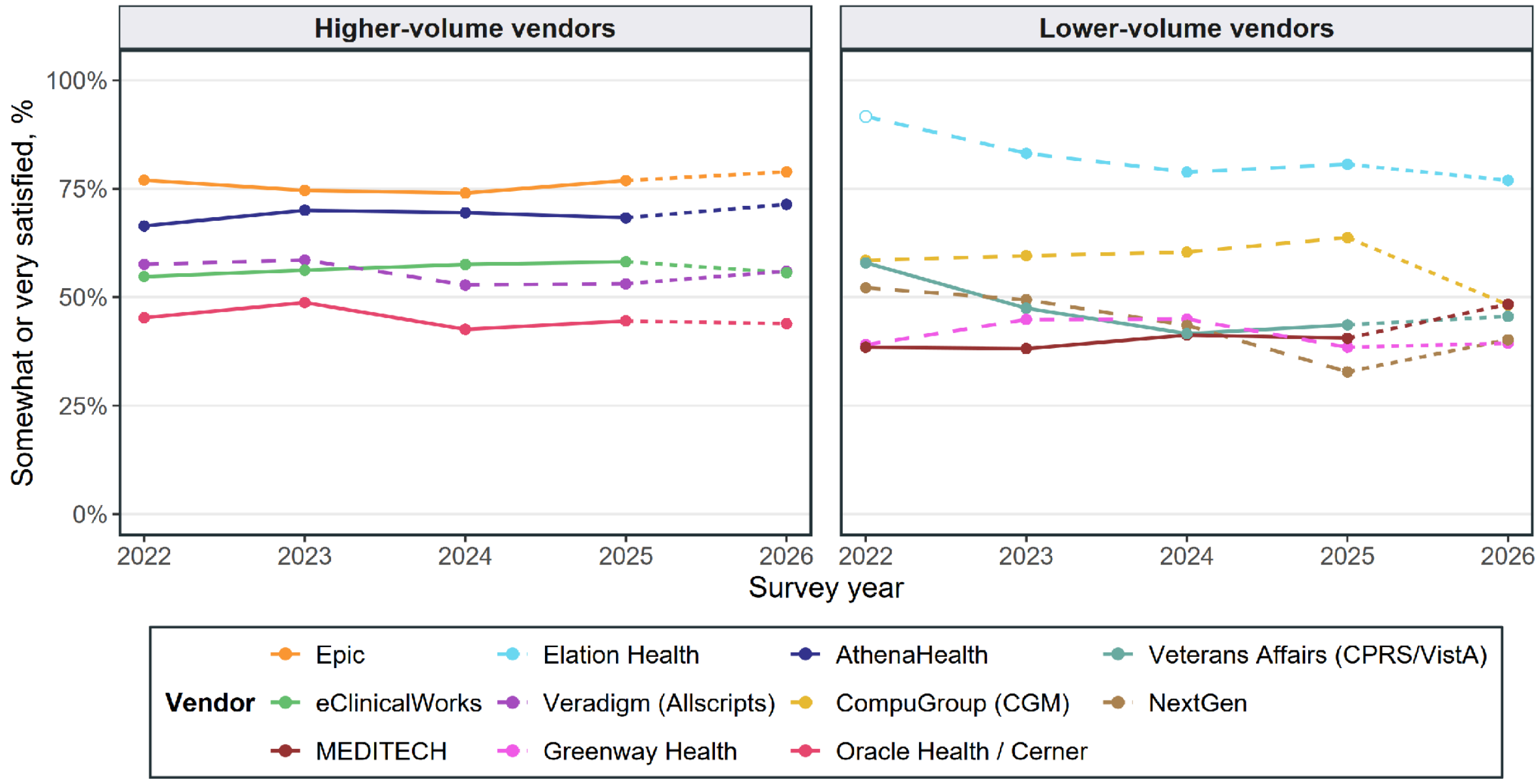

**Figure S1.** Somewhat or very satisfied with the EHR, by vendor and survey year, 2022–2026 (sensitivity threshold). Higher- and lower-volume vendors are shown in separate panels. Hollow points, n = 15–49. Dashed segment, 2026 wave.

### Figure S2:

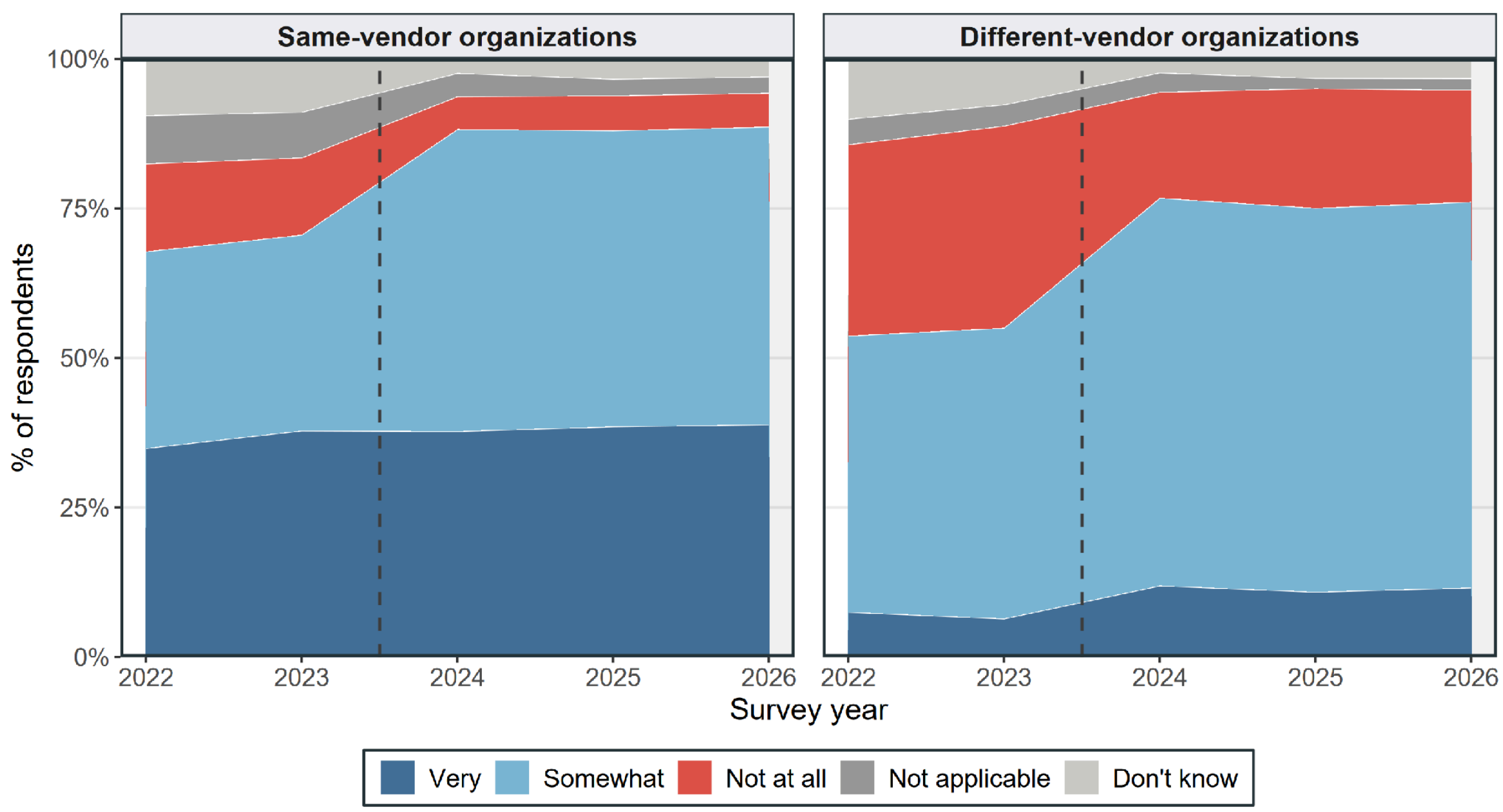

**Figure S2.** Full distribution of responses to the ease-of-using-outside-information items, by exchange type and survey year, including "not applicable" and "don't know." The dashed line marks the 2024 questionnaire revision, after which only respondents reporting the ability to exchange were asked.

**Figure S3A:**

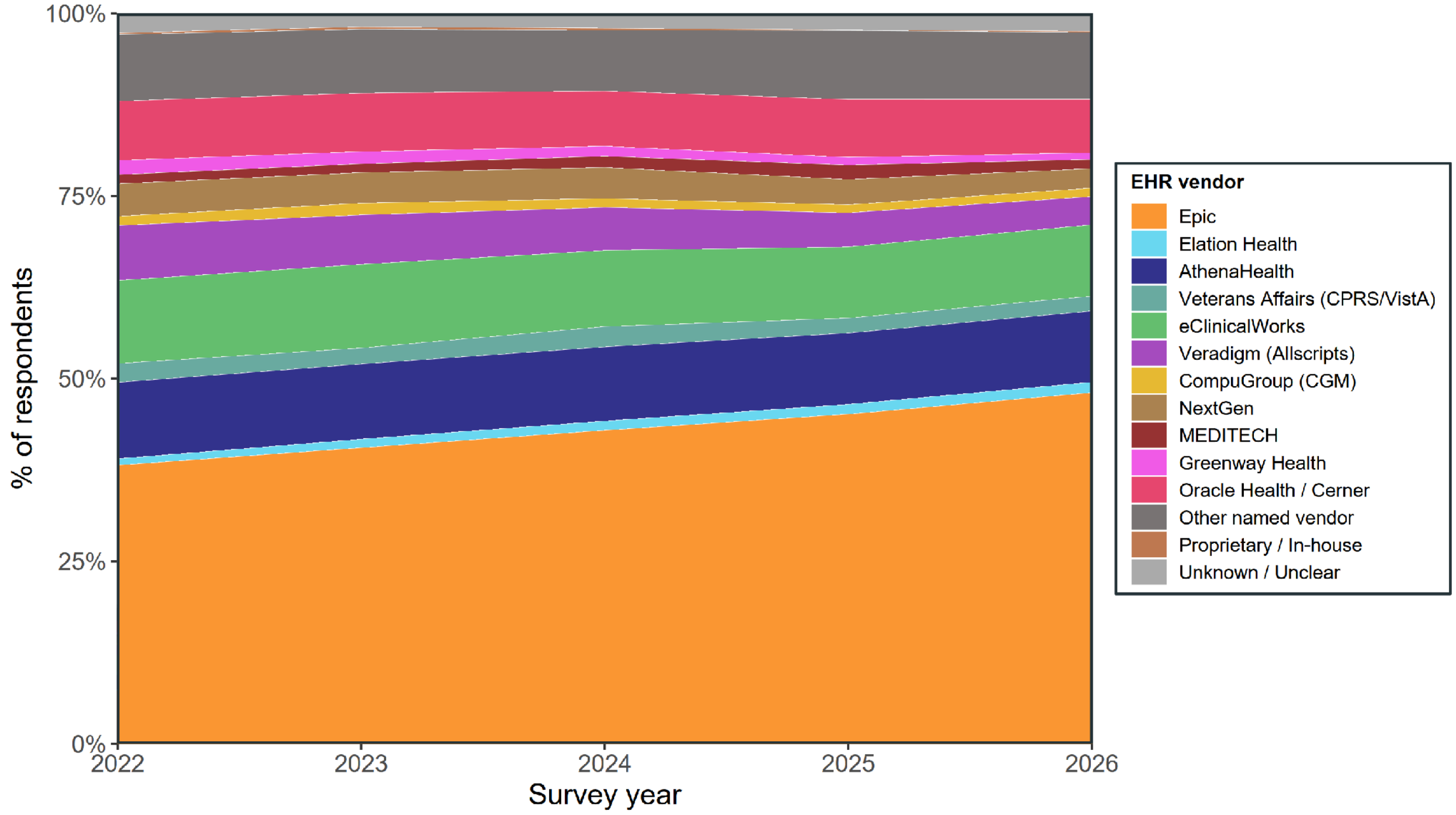

**Figure S3B:**

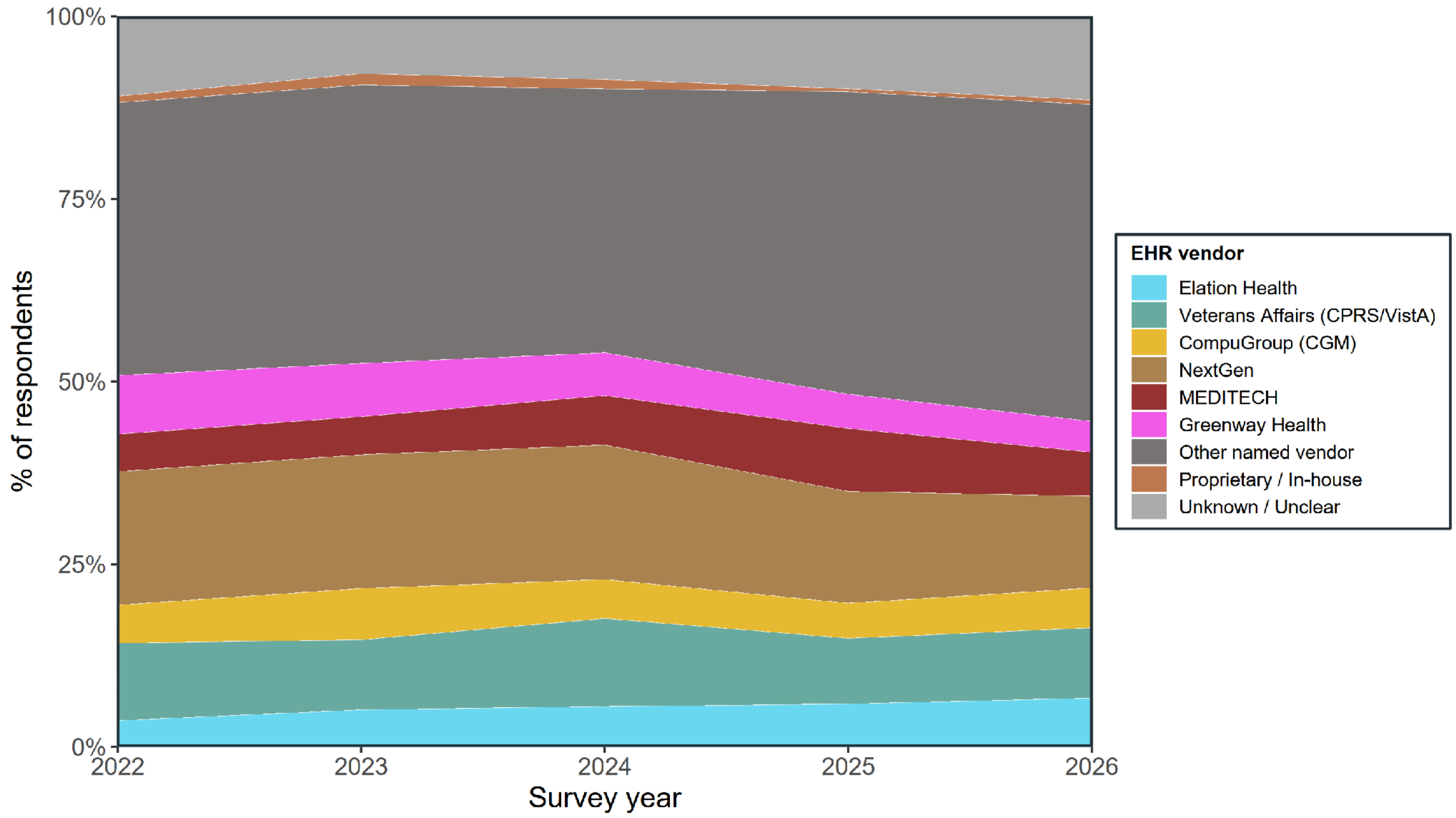

**Figure S3.** EHR vendor composition of respondents by survey year. A, all vendors and residual categories; B, excluding the five highest-volume vendors.

**Figure S4A:**

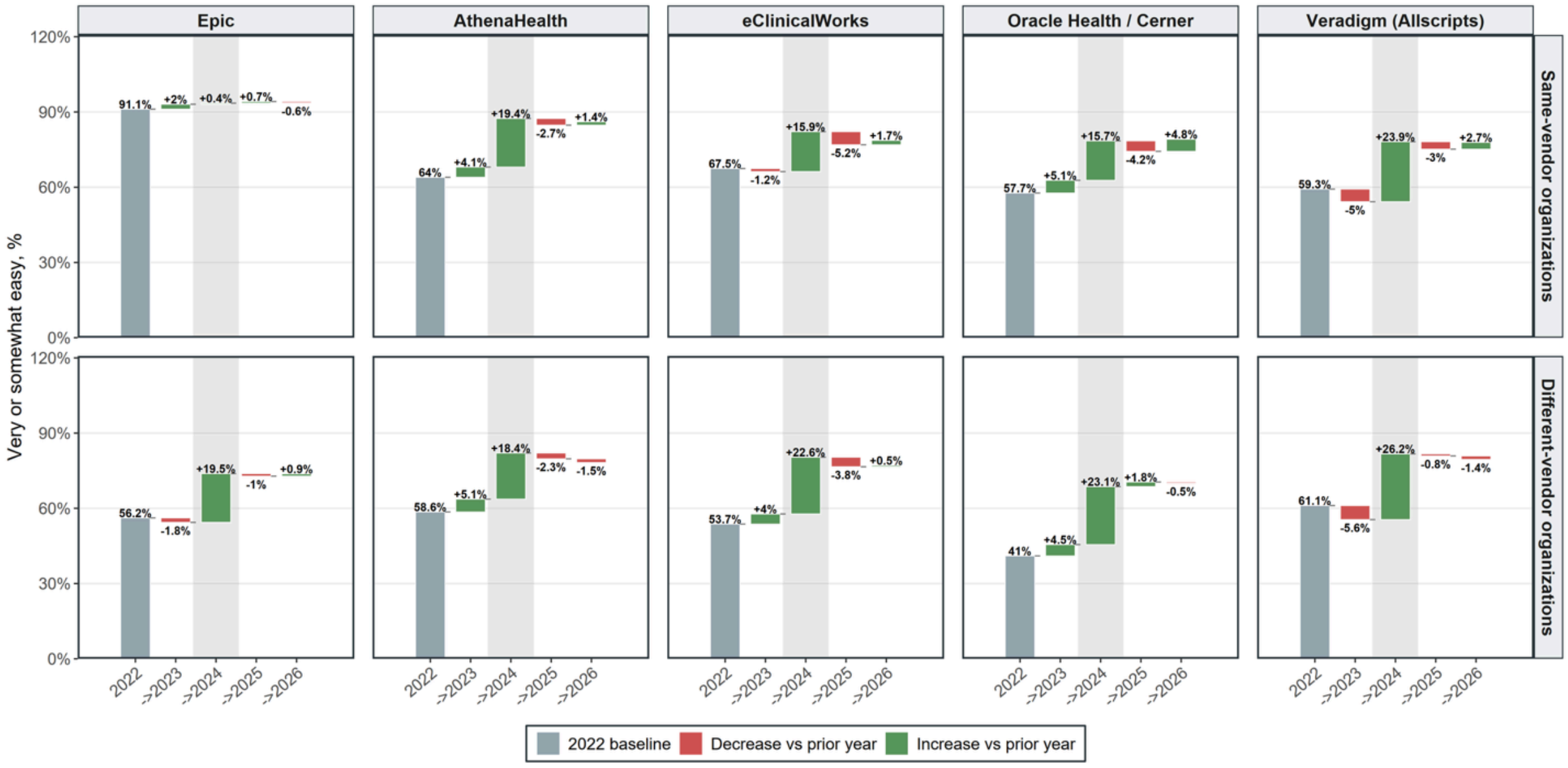

**Figure S4B:**

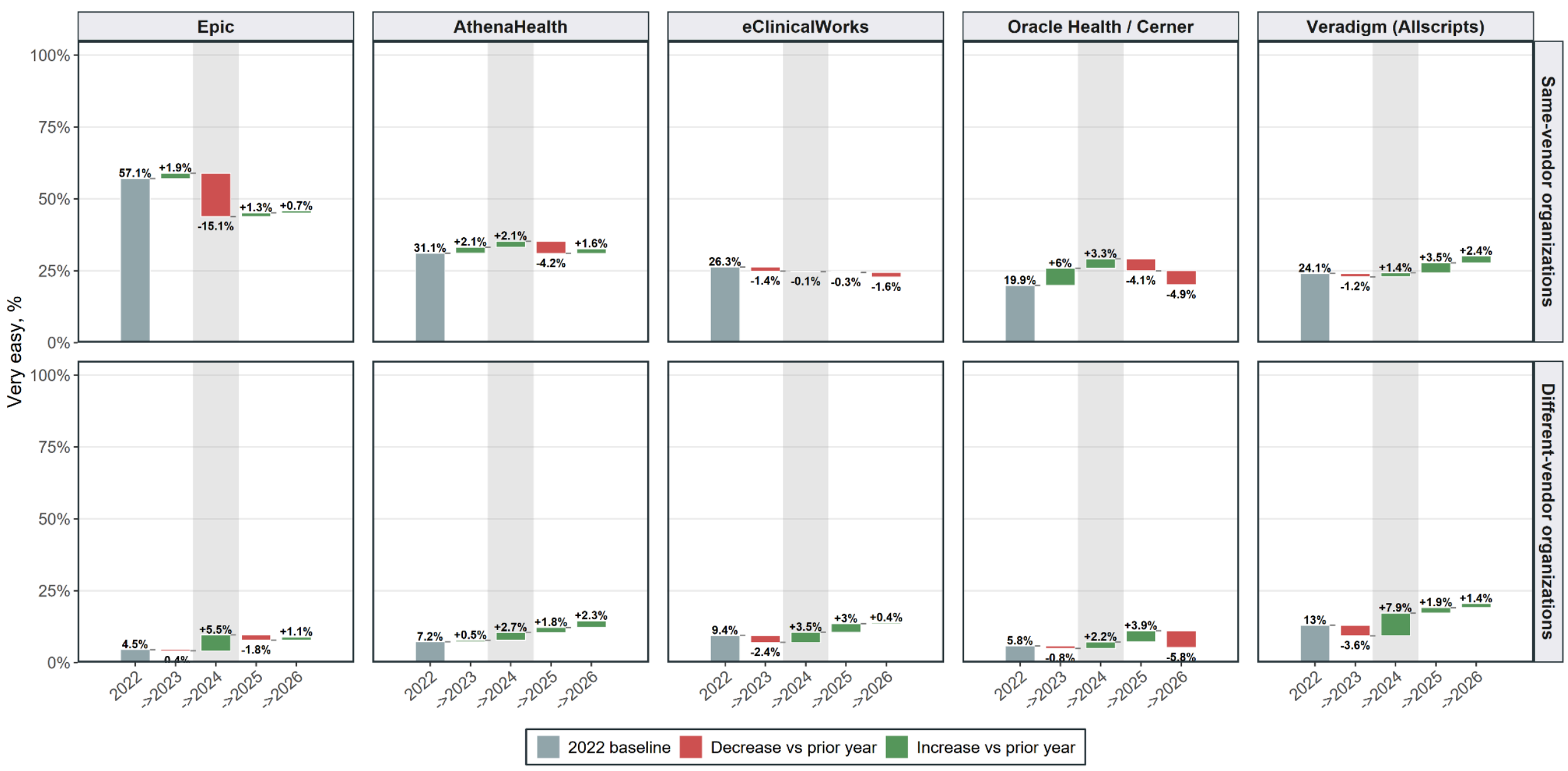

**Figure S4.** Year-over-year change in rating outside information easy to use, higher-volume vendors, by exchange type. A, very or somewhat easy; B, very easy. Gray bars, 2022 baseline; green and red bars, increase or decrease from the prior year. The shaded column spans the 2024 questionnaire revision and is not interpretable as change.

**Figure S5A:**

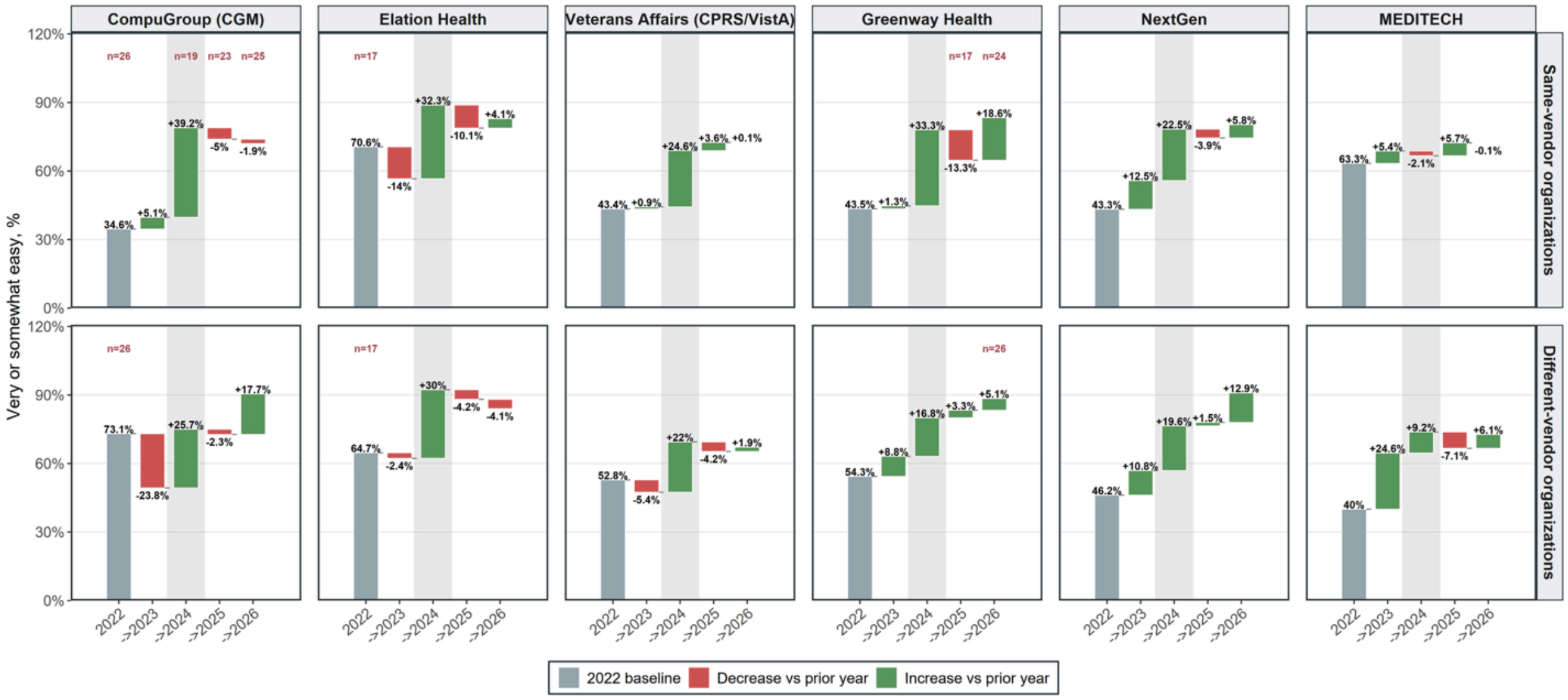

**Figure S5B:**

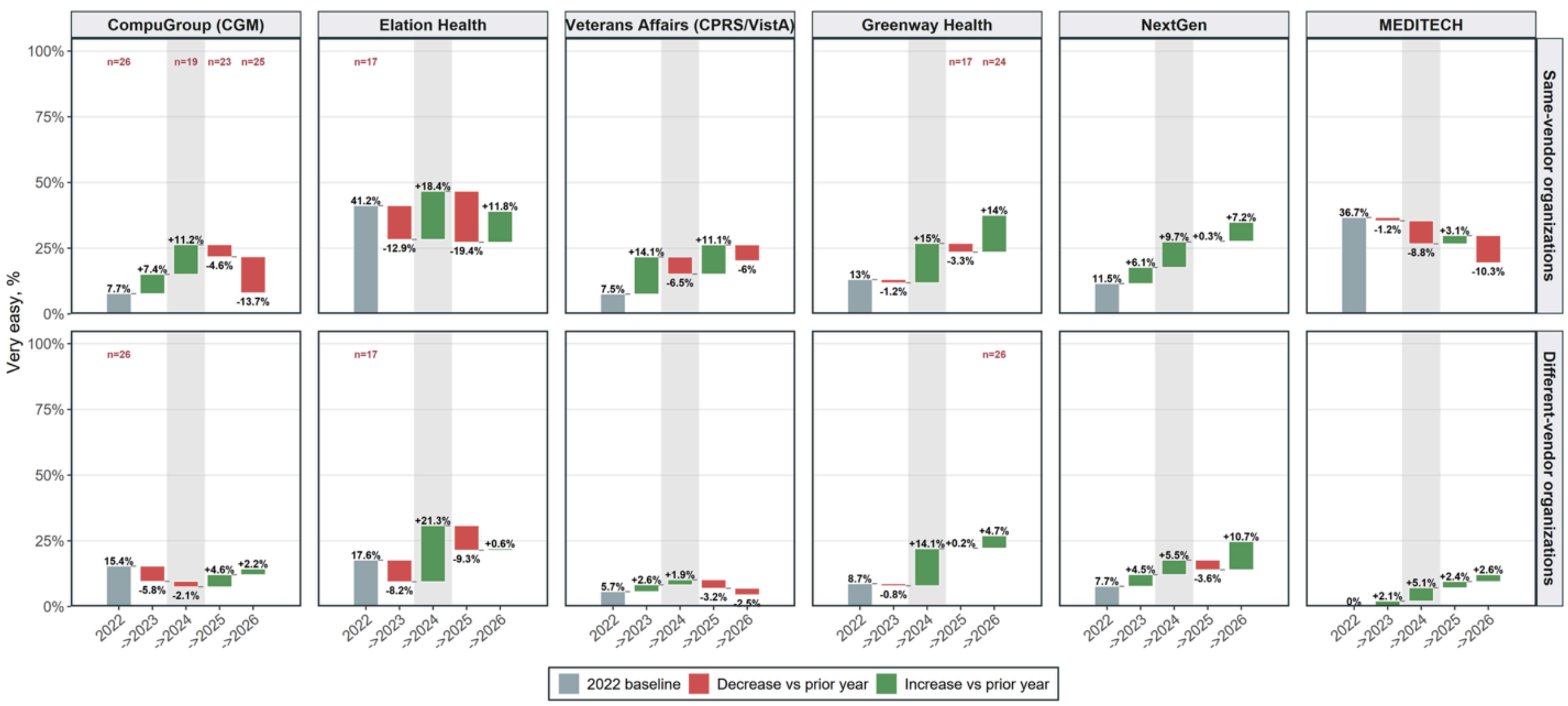

**Figure S5.** As in Figure S4, for lower-volume vendors. Red n labels mark years with fewer than 30 responses.

**Figure S6A,B:**

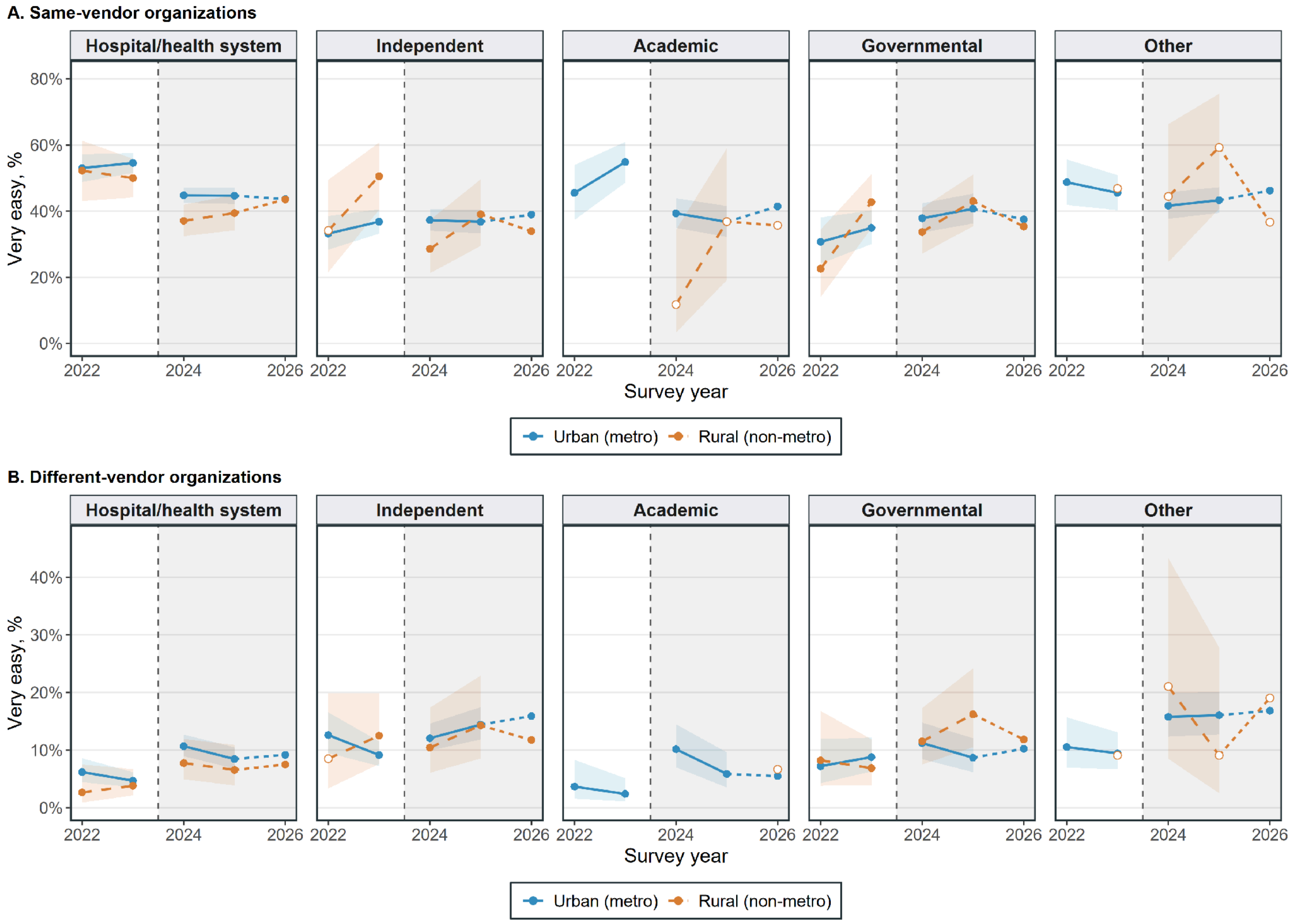

**Figure S6.** Percentage rating outside information very easy, urban versus rural physicians, by practice setting and survey year. A, same-vendor organizations; B, different-vendor organizations. Bands, 95% Wilson CI; shading marks the 2024 questionnaire revision. Hollow points, n = 15–49. Dashed segment, 2026 wave. Panels use separate y-axis ranges.

**Table S1: Vendor Satisfaction Trends and Divergence From Epic, 2022–2026**

| Term | OR per year (95% CI) | P value | Departure from linear trend, P |
|---|---|---|---|
| **A. Own trend per vendor, very satisfied (primary outcome)** | | | |
| Epic | 1.12 (1.09–1.15) | <.001* | <.001* |
| AthenaHealth | 1.02 (0.97–1.08) | .84 | .74 |
| Veterans Affairs (CPRS/VistA) | 0.97 (0.84–1.11) | .84 | .74 |
| eClinicalWorks | 1.05 (0.98–1.12) | .54 | .74 |
| CompuGroup (CGM) | 1.01 (0.84–1.22) | .88 | .74 |
| Veradigm (Allscripts) | 1.02 (0.93–1.12) | .84 | .74 |
| NextGen | 0.84 (0.73–0.96) | .06 | .74 |
| Greenway Health | 1.03 (0.82–1.28) | .88 | .91 |
| Oracle Health / Cerner | 0.96 (0.86–1.06) | .84 | .74 |
| MEDITECH | 1.08 (0.86–1.38) | .84 | .74 |
| **B. Own trend per vendor, top-2 satisfaction (sensitivity)** | | | |
| Epic | 1.05 (1.02–1.08) | .002* | .006* |
| AthenaHealth | 1.03 (0.98–1.09) | .36 | .71 |
| Veterans Affairs (CPRS/VistA) | 0.91 (0.81–1.01) | .24 | .54 |
| eClinicalWorks | 1.01 (0.96–1.06) | .70 | .74 |
| CompuGroup (CGM) | 0.92 (0.80–1.07) | .36 | .57 |
| Veradigm (Allscripts) | 0.96 (0.89–1.03) | .36 | .57 |
| NextGen | 0.84 (0.77–0.91) | <.001* | .37 |
| Greenway Health | 0.97 (0.84–1.12) | .71 | .74 |
| Oracle Health / Cerner | 0.97 (0.91–1.02) | .36 | .57 |
| MEDITECH | 1.10 (0.95–1.27) | .36 | .84 |

| Term | OR per year (95% CI) | P value | Departure from linear trend, P |
|---|---|---|---|
| **C. Trend relative to Epic, unadjusted** | | | |
| Epic (own trend per year) | 1.12 (1.09–1.15) | <.001* | – |
| Elation Health | 0.99 (0.85–1.15) | .88 | – |
| AthenaHealth | 0.91 (0.86–0.97) | .003* | – |
| Veterans Affairs (CPRS/VistA) | 0.87 (0.75–1.00) | .046* | – |
| eClinicalWorks | 0.94 (0.87–1.00) | .06 | – |
| CompuGroup (CGM) | 0.91 (0.75–1.09) | .30 | – |
| Veradigm (Allscripts) | 0.91 (0.83–1.00) | .06 | – |
| NextGen | 0.75 (0.65–0.86) | <.001* | – |
| Greenway Health | 0.92 (0.73–1.14) | .45 | – |
| Oracle Health / Cerner | 0.85 (0.77–0.95) | .003* | – |
| MEDITECH | 0.97 (0.76–1.23) | .79 | – |
| **D. Trend relative to Epic, adjusted** | | | |
| Epic (own trend per year) | 1.12 (1.09–1.15) | <.001* | – |
| Elation Health | 0.97 (0.83–1.13) | .66 | – |
| AthenaHealth | 0.90 (0.85–0.96) | .001* | – |
| Veterans Affairs (CPRS/VistA) | 0.81 (0.62–1.07) | .14 | – |
| eClinicalWorks | 0.93 (0.87–1.00) | .06 | – |
| CompuGroup (CGM) | 0.91 (0.75–1.10) | .33 | – |
| Veradigm (Allscripts) | 0.89 (0.80–0.98) | .02* | – |
| NextGen | 0.79 (0.68–0.92) | .002* | – |
| Greenway Health | 0.94 (0.75–1.18) | .58 | – |
| Oracle Health / Cerner | 0.87 (0.78–0.98) | .02* | – |
| MEDITECH | 0.95 (0.74–1.23) | .71 | – |

**Table 1.** Logistic models with a linear year term. A–B, each vendor's own per-year trend, for vendors with at least 50 respondents in every year; P values Benjamini-Hochberg (BH) adjusted across vendors. C–D, ratio of each vendor's per-year trend to Epic's (vendor × year interaction); P values unadjusted; D is adjusted for the Table S4 covariates. The departure test compares each model with one treating year as a factor. OR, odds ratio; CI, confidence interval; –, not applicable. *P < .05.

**Table S2: Tests for Effect Modification of Ease of Using Outside Information**

| Series | Period | Interaction tested | df | LR chi-square | P value | BH-adjusted P |
|---|---|---|---|---|---|---|
| **A. Within each measurement period (year linear)** | | | | | | |
| Different-vendor | 2022–2023 | Rurality × practice setting | 4 | 2.53 | .64 | .69 |
| Different-vendor | 2022–2023 | Epic vs other vendors × practice setting | 4 | 3.62 | .46 | .69 |
| Different-vendor | 2022–2023 | Epic vs other vendors × rurality | 1 | 3.40 | .07 | .16 |
| Different-vendor | 2024–2026 | Rurality × practice setting | 4 | 6.96 | .14 | .24 |
| Different-vendor | 2024–2026 | Epic vs other vendors × practice setting | 4 | 11.60 | .02 | .08 |
| Different-vendor | 2024–2026 | Epic vs other vendors × rurality | 1 | 3.86 | .05 | .15 |
| Same-vendor | 2022–2023 | Rurality × practice setting | 4 | 7.25 | .12 | .24 |
| Same-vendor | 2022–2023 | Epic vs other vendors × practice setting | 4 | 17.13 | .002 | .01* |
| Same-vendor | 2022–2023 | Epic vs other vendors × rurality | 1 | 0.16 | .69 | .69 |
| Same-vendor | 2024–2026 | Rurality × practice setting | 4 | 3.22 | .52 | .69 |
| Same-vendor | 2024–2026 | Epic vs other vendors × practice setting | 4 | 17.52 | .002 | .01* |
| Same-vendor | 2024–2026 | Epic vs other vendors × rurality | 1 | 0.22 | .64 | .69 |
| **B. Pooled across periods (survey period adjusted)** | | | | | | |
| Different-vendor | 2022–2026 | Rurality × practice setting | 4 | 9.19 | .06 | .08 |
| Different-vendor | 2022–2026 | Epic vs other vendors × practice setting | 4 | 12.73 | .01 | .04* |
| Different-vendor | 2022–2026 | Epic vs other vendors × rurality | 1 | 0.92 | .34 | .34 |
| Same-vendor | 2022–2026 | Rurality × practice setting | 4 | 1.28 | .86 | .90 |
| Same-vendor | 2022–2026 | Epic vs other vendors × practice setting | 4 | 33.14 | <.001 | <.001* |
| Same-vendor | 2022–2026 | Epic vs other vendors × rurality | 1 | 0.01 | .90 | .90 |

**Table 2.** Block likelihood-ratio (LR) tests from logistic models of very easy, fitted separately by exchange type. A, within each measurement period, with a linear year term; B, both periods pooled, adjusted for period. Epic vs other vendors, Epic versus the ten other modeled vendors; df, degrees of freedom. Benjamini-Hochberg (BH) correction within block. *BH-adjusted P < .05.

**Table S3: Contribution of Each Predictor Block to the Adjusted Satisfaction Model**

| Predictor block | df | LR chi-square | Chi-square per df | Change in McFadden $R^2$ | Change in AUC | P value |
|---|---|---|---|---|---|---|
| EHR vendor | 10 | 1947.0 | 194.7 | 0.0551 | +0.0910 | <.001* |
| Site ownership | 4 | 88.4 | 22.1 | 0.0025 | +0.0031 | <.001* |
| Survey year | 4 | 70.0 | 17.5 | 0.0020 | +0.0060 | <.001* |
| Census region | 3 | 43.3 | 14.4 | 0.0012 | +0.0023 | <.001* |
| Practice setting | 4 | 41.9 | 10.5 | 0.0012 | +0.0017 | <.001* |
| Practice size | 3 | 6.6 | 2.2 | 0.0002 | +0.0002 | .09 |
| Rurality | 1 | 0.0 | 0.0 | 0.0000 | +0.0000 | .95 |

**Table 3.** Each row compares the full adjusted model (Table S4) with the same model omitting that block, on the same sample. Changes in McFadden pseudo-$R^2$ and AUC are not inflated by the number of categories. LR, likelihood ratio; df, degrees of freedom; AUC, area under the receiver operating characteristic curve. *$P < .05$.

**Table S4: Adjusted Odds of Being Very Satisfied With the EHR, 2022–2026, Under Alternative Specifications**

| Term | OR (95% CI) | P value | Group | Setting and rurality only | Ordinal model |
|---|---|---|---|---|---|
| **EHR vendor (reference: Epic)** | | | | | |
| Epic (reference) | 1.00 (reference) | – | a | 1.00 (reference) | 1.00 (reference) |
| Elation Health | 0.91 (0.74–1.11) | .35 | b | 1.05 (0.86–1.28) | 0.93 (0.77–1.12) |
| AthenaHealth | 0.64 (0.59–0.70) | <.001* | c | 0.69 (0.63–0.75) | 0.63 (0.59–0.68) |
| Veterans Affairs (CPRS/VistA) | 0.39 (0.30–0.50) | <.001* | d | 0.42 (0.32–0.53) | 0.31 (0.26–0.36) |
| eClinicalWorks | 0.27 (0.25–0.30) | <.001* | d | 0.30 (0.27–0.34) | 0.32 (0.30–0.35) |
| CompuGroup (CGM) | 0.25 (0.19–0.32) | <.001* | d | 0.30 (0.23–0.38) | 0.29 (0.24–0.35) |
| Veradigm (Allscripts) | 0.24 (0.21–0.27) | <.001* | d | 0.27 (0.23–0.30) | 0.30 (0.27–0.32) |
| NextGen | 0.20 (0.17–0.24) | <.001* | e | 0.21 (0.17–0.25) | 0.23 (0.21–0.26) |
| Greenway Health | 0.19 (0.14–0.25) | <.001* | ef | 0.19 (0.14–0.26) | 0.19 (0.16–0.23) |
| Oracle Health / Cerner | 0.17 (0.15–0.20) | <.001* | f | 0.17 (0.15–0.20) | 0.26 (0.24–0.28) |
| MEDITECH | 0.17 (0.12–0.23) | <.001* | ef | 0.17 (0.12–0.23) | 0.22 (0.19–0.26) |
| **Rurality (reference: urban)** | | | | | |
| Rural (non-metro) | 1.00 (0.92–1.08) | .95 | – | 1.00 (0.93–1.08) | 1.01 (0.95–1.07) |
| **Practice setting (reference: hospital/health system)** | | | | | |
| Independent | 0.98 (0.89–1.09) | .78 | – | 1.33 (1.23–1.45) | 1.06 (0.98–1.15) |
| Academic | 0.90 (0.81–0.99) | .04* | – | 0.87 (0.79–0.96) | 0.99 (0.91–1.07) |
| Governmental | 0.99 (0.89–1.10) | .86 | – | 0.95 (0.85–1.06) | 1.02 (0.94–1.10) |
| Other | 1.25 (1.15–1.37) | <.001* | – | 1.31 (1.21–1.42) | 1.12 (1.04–1.20) |

| Term | OR (95% CI) | P value | Group | Setting and rurality only | Ordinal model |
|---|---|---|---|---|---|
| **Site ownership (reference: no ownership stake, 100% employed)** | | | | | |
| Other | 1.12 (0.90–1.39) | .30 | – | – | 1.02 (0.85–1.22) |
| Partial owner or shareholder | 1.25 (1.14–1.37) | <.001* | – | – | 1.24 (1.15–1.33) |
| Self-employed as a contractor (including locums) | 0.96 (0.79–1.15) | .63 | – | – | 1.02 (0.89–1.17) |
| Sole owner | 1.92 (1.67–2.21) | <.001* | – | – | 1.71 (1.53–1.92) |
| **Practice size (reference: >20 clinicians)** | | | | | |
| 2–5 clinicians | 1.08 (1.00–1.17) | .06 | – | – | 1.09 (1.03–1.16) |
| 6–20 clinicians | 1.00 (0.93–1.08) | .91 | – | – | 1.05 (0.99–1.11) |
| Solo practice | 1.13 (0.98–1.30) | .09 | – | – | 1.09 (0.98–1.22) |
| **Census region (reference: Midwest)** | | | | | |
| Northeast | 0.88 (0.81–0.97) | .008* | – | – | 0.86 (0.80–0.92) |
| South | 1.14 (1.07–1.23) | <.001* | – | – | 1.07 (1.01–1.13) |
| West | 0.96 (0.89–1.04) | .35 | – | – | 0.97 (0.91–1.03) |
| **Survey year (reference: 2022)** | | | | | |
| Year 2023 | 0.92 (0.84–1.01) | .07 | – | 0.92 (0.84–1.01) | – |
| Year 2024 | 0.95 (0.87–1.05) | .31 | – | 0.96 (0.88–1.06) | – |
| Year 2025 | 1.12 (1.02–1.24) | .02* | – | 1.13 (1.03–1.25) | – |
| Year 2026 | 1.24 (1.12–1.36) | <.001* | – | 1.25 (1.13–1.38) | – |

**Table 4.** Primary model (OR and P value columns): logistic regression of very satisfied on all listed terms (n = 29,337). Group: vendors sharing a letter did not differ significantly (Benjamini-Hochberg-adjusted pairwise tests of the unadjusted proportions). Setting and rurality only: the same model without ownership, practice size, and region. Ordinal: proportional-odds model over the five-point scale with a linear year term, so survey-year rows are blank. OR, odds ratio; CI, confidence interval; –, not applicable or not in that model. *P < .05.

**Table S5: Within-Period Trends in Ease of Using Outside Information**

| Series | Period | Start, % | End, % | Change, pp | P value | BH-adjusted P |
|---|---|---|---|---|---|---|
| **A. Very easy (primary outcome)** | | | | | | |
| Different-vendor | 2022–2023 | 8.8 | 7.2 | –1.5 | .04 | .09 |
| Different-vendor | 2024–2026 | 12.6 | 12.2 | –0.4 | .60 | .60 |
| Same-vendor | 2022–2023 | 42.3 | 45.3 | +3.0 | .04 | .09 |
| Same-vendor | 2024–2026 | 40.3 | 41.2 | +0.9 | .34 | .46 |
| **B. Very or somewhat easy (sensitivity)** | | | | | | |
| Different-vendor | 2022–2023 | 62.7 | 62.0 | –0.7 | .61 | .77 |
| Different-vendor | 2024–2026 | 81.3 | 80.3 | –1.0 | .27 | .55 |
| Same-vendor | 2022–2023 | 82.2 | 84.5 | +2.3 | .03 | .13 |
| Same-vendor | 2024–2026 | 94.1 | 94.0 | –0.1 | .77 | .77 |

**Table 5.** Cochran-Armitage tests for trend within each measurement period; Benjamini-Hochberg (BH) correction across the four tests at each threshold. A, very easy; B, very or somewhat easy. Change, pp, percentage-point change from the first to the last year of the period. *BH-adjusted P < .05.

**Table S6: Within-Period Trends in Ease of Using Outside Information, by Vendor**

| Vendor | Series | Period | Start, % | End, % | Change | P | BH-adj |
|---|---|---|---|---|---|---|---|
| **A. Very easy (primary outcome)** | | | | | | | |
| Epic | Same | 2022–2023 | 60.1 | 61.6 | +1.6 | .47 | .87 |
| Epic | Same | 2024–2026 | 45.1 | 47.1 | +2.0 | .10 | .35 |
| AthenaHealth | Same | 2022–2023 | 39.4 | 41.4 | +2.0 | .67 | .87 |
| AthenaHealth | Same | 2024–2026 | 38.3 | 35.2 | –3.1 | .33 | .39 |
| Veterans Affairs | Same | 2022–2023 | 12.1 | 31.8 | +19.7 | .03 | .27 |
| Veterans Affairs | Same | 2024–2026 | 18.5 | 25.0 | +6.5 | .24 | .35 |
| eClinicalWorks | Same | 2022–2023 | 31.6 | 30.1 | –1.5 | .70 | .87 |
| eClinicalWorks | Same | 2024–2026 | 27.2 | 25.4 | –1.8 | .58 | .58 |
| Veradigm | Same | 2022–2023 | 29.8 | 28.4 | –1.3 | .79 | .87 |
| Veradigm | Same | 2024–2026 | 27.9 | 35.1 | +7.1 | .20 | .35 |
| NextGen | Same | 2022–2023 | 16.4 | 22.1 | +5.6 | .33 | .87 |
| NextGen | Same | 2024–2026 | 30.3 | 39.0 | +8.7 | .25 | .35 |
| Greenway | Same | 2022–2023 | 16.7 | 18.0 | +1.3 | .87 | .87 |
| Oracle / Cerner | Same | 2022–2023 | 24.0 | 31.2 | +7.2 | .14 | .55 |
| Oracle / Cerner | Same | 2024–2026 | 32.7 | 21.5 | –11.2 | .003 | .02* |
| Epic | Different | 2022–2023 | 5.0 | 4.4 | –0.5 | .57 | .95 |
| Epic | Different | 2024–2026 | 10.1 | 9.2 | –0.9 | .39 | .42 |
| Elation Health | Different | 2024–2026 | 30.8 | 22.4 | –8.3 | .39 | .42 |
| AthenaHealth | Different | 2022–2023 | 8.6 | 8.8 | +0.2 | .95 | .95 |
| AthenaHealth | Different | 2024–2026 | 10.9 | 15.2 | +4.4 | .06 | .42 |
| Veterans Affairs | Different | 2022–2023 | 7.5 | 10.1 | +2.6 | .64 | .95 |
| Veterans Affairs | Different | 2024–2026 | 11.0 | 4.9 | –6.1 | .16 | .42 |
| eClinicalWorks | Different | 2022–2023 | 11.3 | 7.9 | –3.4 | .16 | .93 |
| eClinicalWorks | Different | 2024–2026 | 11.0 | 14.8 | +3.8 | .12 | .42 |
| Veradigm | Different | 2022–2023 | 14.4 | 10.6 | –3.8 | .28 | .93 |
| Veradigm | Different | 2024–2026 | 18.3 | 22.6 | +4.4 | .38 | .42 |
| CompuGroup | Different | 2024–2026 | 8.3 | 14.6 | +6.3 | .40 | .42 |
| NextGen | Different | 2022–2023 | 9.5 | 13.7 | +4.1 | .35 | .93 |
| NextGen | Different | 2024–2026 | 19.0 | 25.3 | +6.3 | .38 | .42 |
| Greenway | Different | 2022–2023 | 9.8 | 8.8 | –0.9 | .87 | .95 |
| Oracle / Cerner | Different | 2022–2023 | 6.5 | 5.6 | –0.9 | .72 | .95 |
| Oracle / Cerner | Different | 2024–2026 | 7.7 | 5.5 | –2.2 | .42 | .42 |
| **B. Very or somewhat easy (sensitivity)** | | | | | | | |
| Epic | Same | 2022–2023 | 95.9 | 97.4 | +1.4 | .06 | .37 |
| Epic | Same | 2024–2026 | 96.2 | 96.4 | +0.1 | .78 | .86 |
| AthenaHealth | Same | 2022–2023 | 81.1 | 84.9 | +3.8 | .28 | .37 |
| AthenaHealth | Same | 2024–2026 | 94.9 | 92.8 | –2.1 | .21 | .54 |

| Vendor | Series | Period | Start, % | End, % | Change | P | BH-adj |
|---|---|---|---|---|---|---|---|
| Veterans Affairs | Same | 2022–2023 | 69.7 | 65.2 | –4.5 | .65 | .74 |
| Veterans Affairs | Same | 2024–2026 | 84.3 | 89.7 | +5.4 | .33 | .57 |
| eClinicalWorks | Same | 2022–2023 | 81.1 | 80.2 | –0.9 | .78 | .78 |
| eClinicalWorks | Same | 2024–2026 | 90.4 | 87.7 | –2.7 | .23 | .54 |
| Veradigm | Same | 2022–2023 | 73.3 | 67.4 | –5.9 | .26 | .37 |
| Veradigm | Same | 2024–2026 | 89.9 | 90.7 | +0.8 | .86 | .86 |
| NextGen | Same | 2022–2023 | 61.6 | 69.7 | +8.0 | .24 | .37 |
| NextGen | Same | 2024–2026 | 86.6 | 89.8 | +3.2 | .54 | .76 |
| Greenway | Same | 2022–2023 | 55.6 | 68.0 | +12.4 | .24 | .37 |
| Oracle / Cerner | Same | 2022–2023 | 69.8 | 75.8 | +6.1 | .20 | .37 |
| Oracle / Cerner | Same | 2024–2026 | 88.3 | 84.5 | –3.7 | .21 | .54 |
| Epic | Different | 2022–2023 | 62.1 | 59.1 | –3.0 | .18 | .46 |
| Epic | Different | 2024–2026 | 77.7 | 76.9 | –0.8 | .58 | .87 |
| Elation Health | Different | 2024–2026 | 92.3 | 85.7 | –6.6 | .32 | .71 |
| AthenaHealth | Different | 2022–2023 | 70.3 | 72.5 | +2.2 | .59 | .68 |
| AthenaHealth | Different | 2024–2026 | 85.4 | 82.2 | –3.3 | .19 | .67 |
| Veterans Affairs | Different | 2022–2023 | 70.0 | 58.2 | –11.8 | .21 | .46 |
| Veterans Affairs | Different | 2024–2026 | 75.0 | 73.8 | –1.2 | .76 | .93 |
| eClinicalWorks | Different | 2022–2023 | 64.6 | 65.2 | +0.6 | .88 | .88 |
| eClinicalWorks | Different | 2024–2026 | 84.2 | 82.3 | –1.9 | .48 | .86 |
| Veradigm | Different | 2022–2023 | 67.8 | 63.0 | –4.8 | .35 | .46 |
| Veradigm | Different | 2024–2026 | 86.9 | 87.7 | +0.9 | .89 | .93 |
| CompuGroup | Different | 2024–2026 | 83.3 | 92.7 | +9.3 | .22 | .67 |
| NextGen | Different | 2022–2023 | 57.1 | 64.0 | +6.8 | .30 | .46 |
| NextGen | Different | 2024–2026 | 82.3 | 93.3 | +11.1 | .04 | .33 |
| Greenway | Different | 2022–2023 | 61.0 | 70.6 | +9.6 | .30 | .46 |
| Oracle / Cerner | Different | 2022–2023 | 46.0 | 51.7 | +5.7 | .27 | .46 |
| Oracle / Cerner | Different | 2024–2026 | 73.9 | 73.5 | –0.4 | .93 | .93 |

**Table 6.** Cochran-Armitage tests run separately for each vendor within each measurement period, at each threshold, for vendors with at least 30 responses in every year of that period. Same, same-vendor organizations; Different, different-vendor organizations; Change, percentage-point change from the first to the last year of the period; P, unadjusted P value; BH-adj, Benjamini-Hochberg adjusted P value (within series, period, and threshold). Veterans Affairs, Veterans Affairs (CPRS/VistA); Veradigm, Veradigm (Allscripts); CompuGroup, CompuGroup (CGM); Greenway, Greenway Health; Oracle / Cerner, Oracle Health / Cerner. *BH-adjusted P < .05.

**Table S7: Change in Reported Ease at the 2024 Questionnaire Revision**

| Threshold | Series | Period OR (95% CI) | P value |
|---|---|---|---|
| Very easy | Different-vendor | 1.64 (1.47–1.84) | <.001* |
| Very easy | Same-vendor | 0.87 (0.81–0.92) | <.001* |
| Very or somewhat easy | Different-vendor | 2.46 (2.29–2.65) | <.001* |
| Very or somewhat easy | Same-vendor | 3.03 (2.74–3.35) | <.001* |

**Table 7.** Odds of reporting easy use in 2024–2026 versus 2022–2023 at each threshold. These estimates size the effect of the 2024 questionnaire revision and do not represent change in experience. OR, odds ratio; CI, confidence interval. *P < .05.

**Table S8: Tests for Differential Satisfaction Trajectories by Practice Setting and Rurality**

| Term | Estimate | P value |
|---|---|---|
| **A. Block likelihood-ratio tests of the interaction** | | |
| Survey year × rurality | LR chi-square 2.05 (df 1) | .15 |
| Survey year × practice setting | LR chi-square 11.95 (df 4) | .02* |
| **B. Per-year trend in very satisfied, by group (95% CI)** | | |
| Hospital or health system, urban (reference): OR per year | 1.10 (1.06–1.13) | <.001* |
| Rural (non-metro): ratio of per-year trends | 1.04 (0.98–1.10) | .15 |
| Independent: ratio of per-year trends | 0.95 (0.90–1.00) | .04* |
| Academic: ratio of per-year trends | 1.02 (0.94–1.10) | .64 |
| Governmental: ratio of per-year trends | 0.92 (0.87–0.98) | .01* |
| Other: ratio of per-year trends | 0.94 (0.88–0.99) | .03* |

**Table 8.** Logistic model of very satisfied with a linear year term, year × practice setting and year × rurality interactions, and vendor. A, block likelihood-ratio (LR) tests; B, per-year odds ratio (OR) in the reference group (hospital or health system, urban) and each group's ratio to it (values below 1 indicate slower improvement). df, degrees of freedom; CI, confidence interval. *P < .05.